\documentclass[structabstract]{aa}  

\usepackage{graphicx}
\usepackage{longtable}
\usepackage{lscape}
\usepackage{amssymb}
\usepackage{endnotes} 
\usepackage{txfonts}
\usepackage{natbib}
\bibpunct{(}{)}{;}{a}{,}{,}

\usepackage{subfig}
\usepackage{setspace}

\begin{document}

   \title{A Study of starless dark cloud LDN 1570: Distance, Dust properties and Magnetic field geometry}
   \titlerunning{Optical polarimetry and photometry of stars towards L1570}

   \author{Eswaraiah, C.\inst{1}
   		  ,
          Maheswar, G.\inst{1,2}
          ,
          Pandey, A. K.\inst{1}
          ,
          Jose, J.\inst{3}
          ,
          Ramaprakash, A. N.\inst{4}
          ,
          Bhatt, H. C. \inst{3}
          }

   \institute{
   Aryabhatta Research Institute of Observational Sciences, Manora Peak, Nainital 263 129, India\\
   Korea Astronomy and Space Science Institute, 61-1, Hwaam-dong, Yuseong-gu, Daejeon 305-348, Republic of Korea\\
   Indian Institute of Astrophysics, II Block, Koramangala, Bangalore 560 034, India\\
   Inter-University Centre for Astronomy and Astrophysics, Ganeshkhind, Pune 411007, India
             }
   \date{Received --- / Accepted ---}

   \abstract
   {}
   {We wish to map the magnetic field geometry and to study the dust properties of the starless cloud, L1570, using multi-wavelength optical polarimetry and photometry of the stars projected on the cloud.}
   {The direction of the magnetic field component parallel to the plane of the sky of a cloud can be obtained using polarimetry of the stars projected on and located behind the cloud. It is believed that the unpolarized light from the stars background to the cloud undergoes selective extinction while passing through non-spherical dust grains that are aligned with their minor axes parallel to the cloud magnetic field. The emerging light becomes partially plane polarized. The  observed polarization vectors trace the direction of the projected magnetic field of the cloud. We made R-band imaging polarimetry of the stars projected on a cloud, L1570, to trace the magnetic field orientation. We also made multi-wavelength polarimetric and photometric observations to constrain the properties of dust in L1570.}
   {We estimated a distance of $394\pm70$ pc to the cloud using 2MASS $JHK_{s}$ colours. Using the values of the Serkowski parameters namely $\sigma_{1}$, $\overline\epsilon$, $\lambda_{max}$ and the position of the stars on near infrared color-color diagram, we identified 13 stars that could possibly have intrinsic polarization and/or rotation in their polarization angles. One star, 2MASS J06075075+1934177, which is a B4Ve spectral type, show the presence of diffuse interstellar bands in the spectrum apart from showing H$\alpha$ line in emission. 
There is an indication for the presence of slightly bigger dust grains towards L1570 on the basis of the dust grain size-indicators such as $\lambda_{max}$ and $R_{V}$ values. 
The magnetic field lines are found to be parallel to the cloud structures seen in the 250$\mu$m images 
(also in 8 $\mu$m and 12 $\mu$m shadow images) of L1570. Based on the magnetic field geometry, the cloud structure and the complex velocity structure, we believe that L1570 is in the process of formation due to the converging flow material mediated by the magnetic field lines. Structure function analysis showed that in the L1570 cloud region the large scale magnetic fields are stronger when compared with the turbulent component of magnetic fields. The estimated magnetic field strengths 
suggest that the L1570 cloud region is sub-critical and hence could be strongly supported by the magnetic field lines.
}
   {}

   \authorrunning{Eswaraiah et al.}
   \maketitle


\section{Introduction}

It has now been recognized that magnetic field plays an important, and perhaps crucial, role in the formation and evolution of molecular clouds and in the star formation process \citep[e.g.,][]{1976ApJ...210..326M, 2000ApJ...540L.103B, 2008A&A...477....9H}. The magnetic field geometry in the outer regions of dark cloud complexes and clouds that are relatively isolated has been mapped by measuring linear polarization of background stars in optical  wavelengths  \citep[e.g.,][]{1976AJ.....81..958V, 1977AJ.....82..198V, 1981ApJ...243..489V, 1985MNRAS.215..275J, 1986AJ.....92..633V, 1986ApJ...309..619M, 1987ApJ...321..855H, 1988AJ.....96..680V, 1990ApJ...359..363G, 1992MNRAS.257...57B, 1993MNRAS.265....1A, 1993A&A...276..507B, 1995ApJ...443L..49A, 1995ApJ...445..269K, 1996MNRAS.279.1191S, 1998MNRAS.300..497R, 1997ApJ...476..717G, 1999MNRAS.308...40B, 1999A&A...349..912H, 2000A&AS..141..175S, 2004MNRAS.348...83B, 2005AIPC..784..736H, 2008A&A...486L..13A, 2010ApJ...723..146F}. It is believed that the light from the stars reddened by aspherical dust grains that are aligned to a cloud magnetic field is partially plane polarized (typically at the level of few per cent) due to dichroic extinction. Although earlier it is believed that Davis-Greenstein \citep*{1951ApJ...114..206D} mechanism could explain the dust grain alignment in the diffuse ISM, the pursuit of developing a successful theory to explain the possible mechanism of the alignment of dust grains with their minor axis parallel to the local magnetic field is still in progress \citep{2003JQSRT..79..881L, 2004ASPC..309..467R}. Regardless of the details of the alignment mechanism, the dichroic or selective  extinction due to aligned, aspherical dust grains would make the polarization vectors to trace the direction of the plane-of-the-sky magnetic field of a cloud. 

A study of projected magnetic field geometry of the molecular clouds in relation with their other properties, like the structure, kinematics, and alignment of any bipolar outflows that may be present in the cloud, could provide us important insight into the role played by the magnetic field in shaping the structure and the dynamics of these objects. But magnetic field maps of a large number of relatively isolated and structurally simple dark clouds that are at different evolutionary stages are required to make statistically reliable studies \citep[e.g.,][]{2009ApJ...704..891L, 2009MNRAS.398..394W}.

The wavelength dependence of polarization towards many galactic directions follows the empirical relation \citep{1974AJ.....79..581C, 1975ApJ...196..261S, 1982AJ.....87..695W} 
\begin{equation}
 P_{\lambda} = P_{max}~\exp[-K ~$ln$^{2} (\lambda_{max}/\lambda)]
\end{equation}
where $P_{\lambda}$ is the percentage polarization at wavelength $\lambda$ and $P_{max}$ is the peak polarization, occurring at wavelength $\lambda_{max}$. The $\lambda_{max}$  is a function of the optical properties and characteristic particle size distribution of aligned grains \citep{1975ApJ...196..261S, 1978ApJ...225..880M}. The value of $P_{max}$ is determined by the column density, the chemical composition, size, shape, and the alignment efficiency of the dust grains. The parameter $K$, an inverse measure of the width of the polarization curve, was treated as a constant by ~\citet*{1975ApJ...196..261S}, who adopted a value of 1.15. The Serkowski relation with $K$=1.15, provides an adequate representation of the observations of interstellar polarization between wavelengths 0.36 and 1.0 $\mu$m. Multi-wavelength polarimetric observations of background stars projected on a molecular cloud, therefore, could provide us useful information regarding the size distribution of dust grains located there. 

Isolated, small Bok globules \citep{1947ApJ...105..255B} are the simplest subset of starless and star forming \citep{1992ApJ...385L..21Y} molecular clouds \citep{1988ApJS...68..257C, 1990ApJ...365L..73Y}. Of these, starless Bok globules are of much interest as they form the simplest laboratories  to study the early evolutionary stages that precede core collapse and subsequent star formation \citep{1995ApJ...445..269K}. LDN 1570 \citep[hereafter L1570, ][]{1962ApJS....7....1L} is same as Barnard 227 \citep{1962ApJS....7....1L} and CB44 \citep{1988ApJS...68..257C}. The optical images of this cloud show an opaque, elongated (along the North-South direction) core surrounded by a more diffuse dust structure. \citet{2009ApJ...707..137S} have considered L1570 as a starless core and concluded that the cloud is approaching collapse based on their study using 8 $\mu$m and 24 $\mu$m shadow images obtained by \textit{Spitzer Space Telescope}. The cloud is assumed to be at distances in the range between 400 pc - 600 pc \citep{1995A&AS..113..325H} with the most probable distance accepted for various studies being 400 pc \citep[e.g.,][]{2005ApJ...622..938G, 2009ApJ...707..137S}. 

In this work, we made $R$ band polarimetry of 127 stars projected on L1570 aimed to map the magnetic field geometry of the cloud and multi-wavelength polarimetry and photometry of 57 and 144 stars, respectively, to characterize the dust properties. This paper is organized in the following manner. First we present details of the observations in section \ref{observe}. The results are presented in section \ref{result}. A discussion on the results obtained is presented in section \ref{discuss}. Finally, we summarize the paper with conclusions in section \ref{conclude}.


\section{Observations}\label{observe}
\subsection{Polarimetry}
Polarimetric observations of the field containing L1570 were carried out on ten nights; namely, 23, 24, 25, 26 November 2009; 23, 24, 27, 28 December 2009 and 27, 31 December 2010 using the ARIES Imaging Polarimeter \citep[AIMPOL,][]{2004BASI...32..159R} mounted at the Cassegrain focus of the 1.04-m Sampurnanand telescope (ST) of the Aryabhatta Research Institute of Observational sciences (ARIES), Manora Peak, India. We used TK 1024$\times$1024 pixel$^2$ CCD camera. The AIMPOL consists of a half-wave plate modulator and a Wollaston prism beam-splitter. The observations were carried out in $B$, $V$, $R_{c}$ and $I_{c}$ ($\lambda_{B_{eff}}$=0.440$\mu$m, $\lambda_{V_{eff}}$=0.53$\mu$m, $\lambda_{Rc_{eff}}$=0.67$\mu$m and $\lambda_{I_{eff}}$=0.80$\mu$m) photometric bands. A total of 10 sub-regions were observed to cover the entire cloud region. In addition to these we made multi-band  polarimetric observations of five sub-regions (four with $V(RI)_{c}$-band and one with $BV(RI)_{c}$-band) projected on the cloud. Each pixel of the CCD corresponds to 1.73 arc sec and the field-of-view (FOV) is $\sim$~8 arc min in diameter on the sky. The FWHM of the stellar images vary from 2 to 3 pixel. The read out noise and the gain of the CCD are 7.0 $e^{-}$  and 11.98 $e^{-}$/ADU respectively. Since AIMPOL is not equipped with a grid, care was taken to exclude the stars that have contaminations from the overlap of ordinary and extraordinary images of one star on the same of another star in the FOV. The data reduction and the procedures followed to measure the polarization of stars are similar to those described in \citet{2011MNRAS.411.1418E, 2012MNRAS.419.2587E}. 

Additional three fields containing L1570 centered around $\alpha_{2000}$=06$^h$ 07$^m$ 48$\fs$951, $\delta_{2000}$=+19$\degr$ 34$\arcmin$ 34$\farcs$25 (in $BVRI$-bands on 12 December 2010); $\alpha_{2000}$=06$^h$ 07$^m$ 18$\fs$551, $\delta_{2000}$=+19$\degr$ 29$\arcmin$ 51$\farcs$18 (in $VRI$-bands on 12 December 2010) and $\alpha_{2000}$=06$^h$ 07$^m$ 02$\fs$057 $\delta_{2000}$=+19$\degr$ 34$\arcmin$ 16$\farcs$37 (in $BVRI$-bands on 13 December 2010) were observed using 2-m telescope of the Inter University Center for Astronomy and Astrophysics (IUCAA) Girawali Observatory, India. The instrument used was the IUCAA Faint Object Spectrograph and Camera (IFOSC) in the polarimetric mode. It employs an EEV 2K $\times$ 2K thinned, back-illuminated CCD with 13.5$\rm{\mu}$m pixels. The gain and the readout noise of the CCD camera are 1.5e$^{-}$/ADU and 4e$^{-}$ respectively. The FOV of the IFOSC in the imaging polarimetric mode is $\sim 4$ arc min in diameter. It measures linear polarization in the wavelength range $0.35-0.85 \mu$m. This instrument also makes use of a Wollaston prism and half-wave plate to observe two orthogonal polarization components that define a Stoke’s parameter. 

To correct the measurements for the instrumental polarization and the zero-point polarization angle, we observed a number of unpolarized and polarized standards, respectively, taken from \cite{1992AJ....104.1563S}. Our measurements for the standard stars are compared with those taken from the \cite{1992AJ....104.1563S} in Table \ref{result_standards}. The observed degree of polarization ($\%$) and position angle ($^\circ$) for the polarized standards are in good agreement, within the observational errors, with those from \cite{1992AJ....104.1563S}. The instrumental polarization of AIMPOL on the 1.04-m ST has been monitored since 2004 for different projects and found to be $\sim0.1\%$ in different bands \citep[e.g.,][]{2004BASI...32..159R, 2007MNRAS.378..881M, 2008MNRAS.388..105M, 2010MNRAS.403.1577M, 2011MNRAS.411.1418E, 2012MNRAS.419.2587E}. No correction for the instrumental polarization was applied to the data obtained using IFOSC since the instrumental polarization of the instrument on the 2-m telescope is found to be \textless 0.05 per cent. 

In short, using AIMPOL and IFOSC we obtained polarimetric observations of 127 stars in single $R$-band, 42 stars in $VRI$-bands and 15 stars in $BVRI$ bands. The results are presented in Tables \ref{r-band-127}, \ref{vri_pol42} and \ref{bvri_pol15} respectively. We observed overlapping fields using AIMPOL and IFOSC purposefully to check the consistency in the results obtained from both the instruments. However, of the data obtained from both the instruments, for overlapping fields, the data obtained from IFOSC was retained because of the better signal-to-noise ratio (S/N).

\subsection{Photometry}

The CCD optical photometric observations of the central region ($\alpha_{2000}$=06$^h$ 07$^m$ 33$\fs$043; $\delta_{2000}$=+19$\degr$ 30$\arcmin$ 52$\farcs$70) of L1570 were carried out in $BVRI$-bands using the 1.04-m ST on 27 November 2010. The 2K$\times$2K CCD with a plate scale of 0.37 arc sec pixel$^{-1}$ covers a FOV of $\sim$ 13$\times$13 arcmin$^{2}$ in the sky. To improve the S/N, the observations were carried out in a binning mode of 2$\times$2 pixels. 
The standard field SA\,98 from \citet{1992AJ....104..340L} was observed on the same night to apply the atmospheric corrections as well as to standardize the observations. SA\,98 was observed at an air mass close to that of L1570 
and the night was photometric with an average seeing of $\sim$2$\arcsec$. We performed point spread function photometry using the {\small DAOPHOT} package in {\small IRAF} on all the processed images to derive the photometric instrumental magnitudes. By using the average extinction coefficients (0.3, 0.2, 0.13, 0.08 for $BVRI$-bands respectively) for the Manora Peak site, we then derived color coefficients using the photometric results of SA\,98. The scattering 
expected in the average extinction coefficient over a period of one year is $\sim$0.05, 0.03, 0.02 and 0.02, respectively for the $B$, $V$, $R$ and $I$ bands in this site. However, since SA\,98 was observed close to the air mass of L1570, the error due to the scattering in the average extinction coefficient can be considered negligible. The central region of L1570 observed with ST was calibrated by applying these extinction and color coefficients. The calibration uncertainties between the standard and transformed $V$ magnitudes and $B-V$, $V-R$, $V-I$ colors were of the order of 0.03 mag. 
 
\subsection{Spectroscopy}
Spectroscopic observations of two of the four emission line stars in the vicinity of L1570 identified in the surveys for H${\alpha}$ emission sources in the northern hemisphere \citep{1983PASJ...35..299O, 1997csnm.book.....K, 1999A&AS..134..255K} were obtained on 15 and 16 October 2010 using the Hanle Faint Object Spectrograph (HFOSC) in the wavelength range from $3800-6840\AA$ with a spectral resolution of  1330. The spectra were obtained to confirm the presence of emission lines in these stars so as to exclude them from the analysis to infer the magnetic field geometry of L1570. All the spectra were bias subtracted, flat-field corrected, extracted and wavelength calibrated in the standard manner using IRAF.

\begin{figure}
\centering
\resizebox{8.5cm}{9.5cm}{\includegraphics{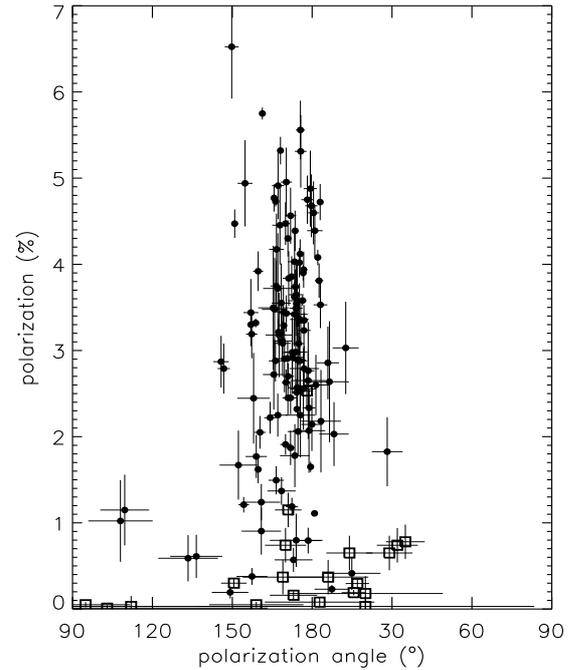}}
\caption{The position angle vs. degree of polarization plot of 127 stars towards L1570 shown using filled circles. The open squares represent stars chosen from a circular area of radius 10$^{\circ}$ about L1570 obtained from \cite{2000AJ....119..923H}.\label{fig:result}}
\end{figure}


\section{Results}\label{result}
\subsection{Polarimetry}
The results obtained from our $R$-band polarization measurements of 127 stars are given in Table \ref{r-band-127}.  The stars are ordered in increasing Right Ascension. The columns of the table give, column 2: Right Ascension ($J2000$); column 3: declination ($J2000$); column 4: degree of polarization in per cent ($P$); column 5: standard error in the degree of polarization in per cent ($\epsilon_{P}$); column 6: position angle in degree ($\theta$); column 7: standard error in the position angle in degree ($\epsilon_{\theta}$); columns 8-13: $JHK{s}$ magnitudes and their corresponding errors obtained from the Two Micron all Sky Survey \citep[2MASS, ][]{2003yCat.2246....0C}. The position angles are measured from the north increasing eastward. We selected only those stars that showed $\epsilon_{P}/P\leqslant0.5$. The observed degree of polarization of the stars range from 0.1 to 6.5 per cent. 

\begin{figure*}
\centering
\resizebox{17cm}{16.5cm}{\includegraphics{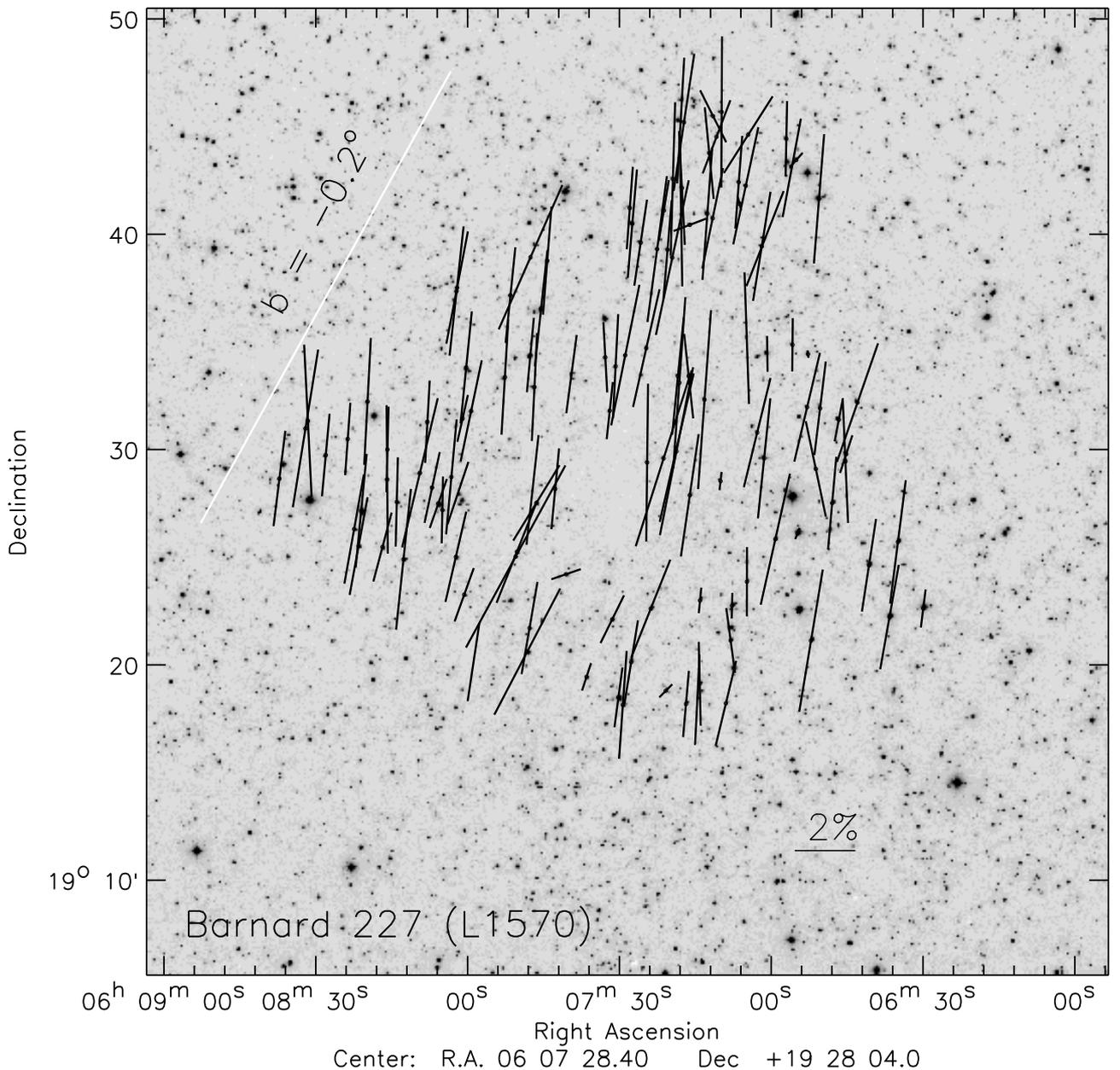}}
\caption{The polarization vectors are over-plotted on the DSS B-band image of the field containing L1570. The length of the vectors corresponds to the degree of polarization and the direction of the orientation corresponds to the polarization position angle of stars measured from the North increasing towards the East.}\label{fig:dss}
\end{figure*}

In Fig. \ref{fig:result}, we present the $P$ ($\%$) versus $ \theta $ ($ ^{o} $) plot for 127 stars observed towards the direction of L1570 (filled circles).  The open squares show stars chosen from a circular area of radius 10$^{\circ}$ about L1570 obtained from the \cite{2000AJ....119..923H} catalogue (see section \ref{sec:foreground} for a description). In Fig. \ref{fig:dss}, we show the R-band polarization vectors of 127 stars overlaid on the B-band image containing L1570 obtained from the Digitized Sky Survey (DSS). The polarization vectors are drawn centered on the stars observed. The length of the vector is proportional to the degree of polarization, $P$ (in per cent), and is oriented in the direction given by the  position angle, ${\theta}$ (in degree). For reference, we have shown a vector corresponding to 2\% polarization with 90$^{\circ}$ orientation. The plane parallel to the Galactic plane at $b=-0.2^{\circ}$, shown using white thick line, projects in this region to a direction 149$^{\circ}$ to the east of the north. The multi-wavelength polarimetric results are presented in Tables \ref{vri_pol42} and \ref{bvri_pol15}. The column 1 gives identification numbers that are same as those given in Table \ref{r-band-127}. Columns 2 and 3 give the Right Ascension ($J2000$) and the  Declination ($J2000$) respectively.

Polarization measurements of 21 stars, mostly lying to the northern parts of L1570, were carried out by \citet{1993A&A...276..507B}. They observed the stars without any filter except in the case of two for which they had observations in multiple filters also. The results on 11 stars that are common in both the studies are compared in Fig. \ref{fig:bhatt_eswar}. Though we can not compare our R-band results with those from \citet{1993A&A...276..507B} directly (because of the difference in the filters used), we find that the results are in agreement within the errors except in the case of few stars. 

\subsection{Photometry}
Results obtained from our $BVRI$ photometry of stars towards L1570 are presented in Table \ref{photdata}. We found a total of 144 stars in the observed field with their photometric errors $\leqslant$ 0.1 mag in $BVRIJHK_{s}$-bands. The $JHK_{s}$ magnitudes of the stars are obtained from the 2MASS. The unique star ids, their magnitudes and corresponding errors in $BVRIJHK_{s}$-bands are tabulated in Table \ref{photdata}. Among these, 29 stars are found to have both polarimetric and photometric data. These stars are identified in Table \ref{photdata} with their corresponding identification number from Table \ref{r-band-127}  (polarimetric results) is shown in parenthesis.

\begin{figure}
\resizebox{8cm}{14cm}{\includegraphics{bhatt_eswar.epsi}}
\caption{The $P\%$ (upper panel) and $\theta$ (lower panel) of stars towards L1570 from \citet{1993A&A...276..507B} (observed without any filter) are compared with those from this work (R-band). The stars marked with their identification numbers taken from Table \ref{r-band-127}. The star \#97 scattered more from the straight line is  found to show (a) H$\alpha$ emission in its spectra (Sec. \ref{spectro}), (b) NIR-excess (Sec. \ref{pol_dust_properties}) and (c) intrinsic 
polarization and polarization angles (Sec. \ref{pol_dust_properties}). \label{fig:bhatt_eswar}}
\end{figure}

\subsection{Spectroscopy}\label{spectro}
\begin{figure*}
\resizebox{18cm}{12cm}{\includegraphics{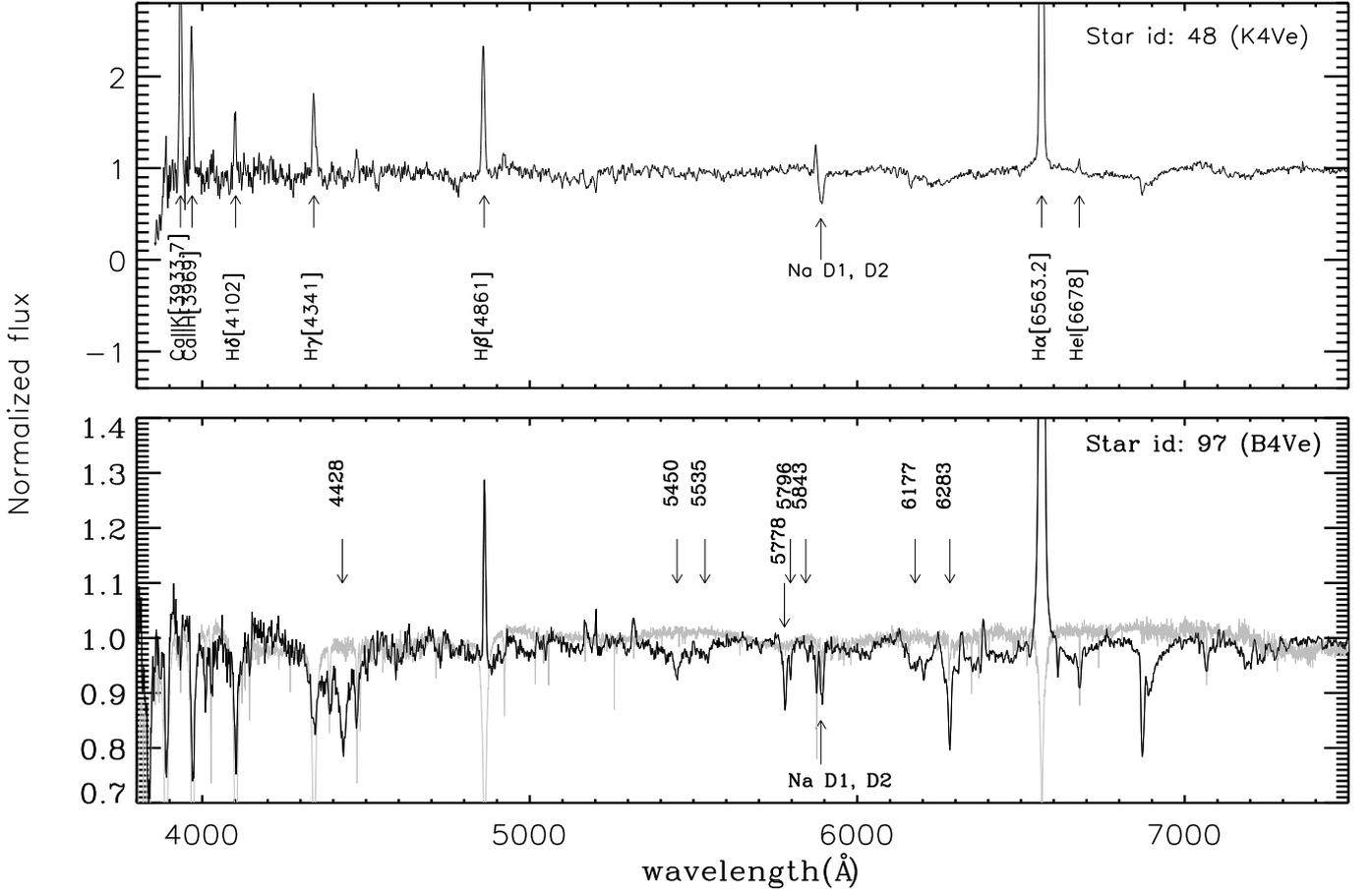}}
\caption{Spectra of the stars  2MASS J06071585+1930001 (star no \# 48, upper panel) and 2MASS J06075075+1934177 (star no \#97, lower panel). These two star show H$\alpha$ in emission. We found a number of DIBs in the spectrum of \#97 that are identified and labeled. The spectrum of a star, HD 189944, with spectral type of B4V and $E(B-V)$=0.065 \citep{1980A&AS...42..251N}, obtained from Indo-US library of coude feed stellar spectra provided by \citet{2004ApJS..152..251V}, is over-plotted, in gray color, on the spectrum of \#97 for a better identification of the DIBs. \label{fig:spec}}
\end{figure*}

Spectra of the two H${\alpha}$ sources, previously recognized by \citet{1983PASJ...35..299O} and by \citet{1997csnm.book.....K,1999A&AS..134..255K}, are shown in Fig. \ref{fig:spec}. These sources are identified with 2MASS J06071585+1930001 (star no \# 48 in Table \ref{r-band-127}, upper panel) and 2MASS J06075075+1934177 (star no \#97, lower panel). These stars are located to the western and to the north-eastern parts of L1570 respectively. We classified these stars as a K4Ve and B4Ve type by comparing their spectrum with those from the stellar library provided by \citet{1998yCat.3092....0J}. The presence of  H$\alpha$ in emission in these two stars are confirmed with an equivalent width of $-44~\AA$ ~and~ $-83~\AA$ for star nos \#48 and \#97 respectively. It is interesting to note that the star \#97 is showing some of the prominent diffused interstellar bands (DIBs) that are identified and labeled in the spectrum with their wavelengths taken from \citet{1995ARA&A..33...19H}. Just for comparison, we over-plotted the spectrum of a star, HD 189944 having a spectral type of B4V and $E(B-V)$=0.065 \citep{1980A&AS...42..251N}, obtained from Indo-US library of coude feed stellar spectra provided by \citet{2004ApJS..152..251V}, using gray color in Fig. \ref{fig:spec}.

\section{Discussion}\label{discuss}

 \subsection{Determination of distance to L1570}\label{sub:dist}
 
In order to subtract interstellar contribution from the observed polarization values, we required to know the distance to L1570. Previous estimates of distances to L1570 were highly uncertain. While \citet{1979PASJ...31..407T}, based on star count method, estimated a distance of 300 pc, \citet{1974AJ.....79...42B} assumed a distance of 400 pc to L1570 based on the number of stars brighter than $m_{pg}=21$ mag projected on the cloud in their photographic plate. Apart from these, no reliable distance estimates are available for L1570 in the literature. Majority of the techniques used to determine distances to molecular clouds are in general extremely tedious and requires considerable amount of telescope time. We used the near-IR photometric method presented by \citet{2010A&A...509A..44M}, which utilizes the vast homogeneous $JHK_{s}$ photometric data produced by the 2MASS that is available for the entire sky, to determine distance to L1570. Colors in the optical wavelengths also could have been used to determine distance but because we have observations of only the central region of L1570, the stars with optical photometry are not sufficient enough for the purpose. 
 
\begin{figure}
\resizebox{8.5cm}{8.5cm}{\includegraphics{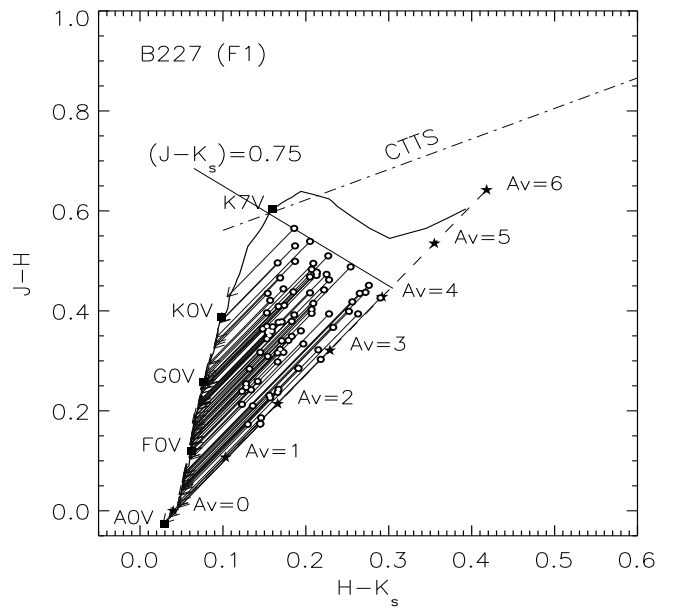}}
\caption{The ($J-H$) vs. ($H-K_{s}$) CC diagram drawn for stars (with $A_{V}\geq1$) from F1 (see Fig. \ref{fig:Av}) region of L1570 to illustrate the method. The solid curve represents locations of  unreddened main sequence stars. The reddening vector for an A0V type star drawn  parallel to the \citet{1985ApJ...288..618R} interstellar reddening vector is shown by the dashed line. The locations of the main sequence stars of different spectral types are marked  with square symbols. The region to the right of the reddening vector is known as the NIR  excess region and corresponds to the location of PMS sources.  The dash-dot-dash line represents the loci of unreddened CTTSs \citep{1997AJ....114..288M}. The open circles represent the observed colors and the arrows are drawn from the observed to the final colors obtained by the method for each star.\label{fig:CC}}
\end{figure}

\begin{figure}
\resizebox{9cm}{6.5cm}{\includegraphics{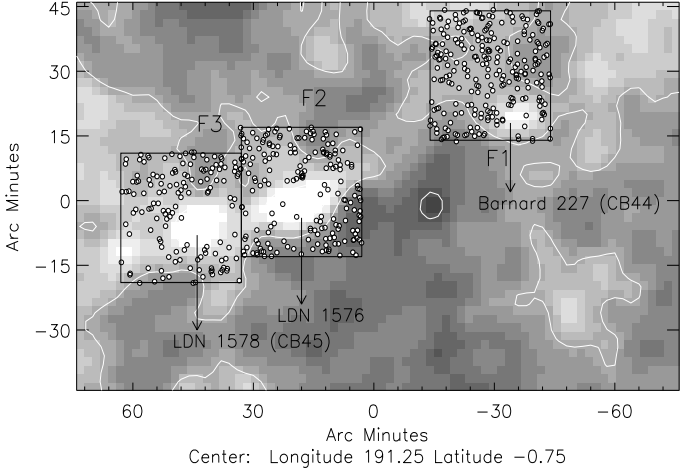}}
\caption{The $2.5^{\circ}\times1.5^{\circ}$ extinction map produced by \citet{2005PASJ...57S...1D} containing L1570 is shown with the fields  F1-F3, each covering $30^{\prime}\times30^{\prime}$ area, marked and labeled. The contours are drawn at 1, 2 and 3 magnitude levels. The stars used for estimating distance to  L1570 are represented by filled circles. The clouds L1570, L1575 and L1578 are identified and labeled.\label{fig:Av}}
\end{figure}

A brief discussion of the method \footnote{A more rigorous discussion on the errors and limitations of the method can be found in \citet{2010A&A...509A..44M}} is presented below. This method is based on a technique that allows spectral classification of stars lying towards the fields containing the clouds into main sequence and giants. In this technique the observed ($J-H$) and ($H-K_{s}$) colors of the stars with ($J-K_{s}$)$\leq0.75$ in ($J-H$) vs. ($H-K_{s}$) color-color (CC) diagram are de-reddened simultaneously using trial values of $A_{V}$  and a normal interstellar extinction law (i.e., total-to-selective extinction value, $R_{V}=3.1$) \footnote{We derived the values of $R_{V}$ towards L1570 using polarimetric and photometric data in section \ref{dust_prop}. Though there are hints of slightly higher value of $R_{V}$ in L1570, we have chosen $R_{V}=3.1$ for the determination of distance because the majority of the stars chosen around L1570 and other cloud regions (see Fig. \ref{fig:Av}) are from the periphery of the clouds and not from the higher extinction inner regions where a higher $R_{V}$ values are usually found \citep[e.g.,][]{2003AJ....126.1888K}.}.
The best fit of the de-reddened colors to the intrinsic  colors giving a minimum value of $\chi^{2}$ then yields the corresponding spectral type and $A_{V}$ for the star. The main sequence stars, thus classified, are then utilized in an $A_{V}$ versus distance plot to bracket the cloud distance. The entire procedure is illustrated in Fig.\ref{fig:CC} where we plot the near infrared CC (NIR-CC) diagram for the stars (with $A_{V}\geq1$) chosen from the region F1 towards the direction of L1570 (see Fig. \ref{fig:Av}). The arrows are drawn from the  observed data points (open circles) to the corresponding de-reddened colors estimated using the method. The maximum extinction values that can be measured using the method are those for A0V type stars ($\approx4$ magnitude).  The extinction traced by stars will fall as we move towards more late type stars.

The giants located relatively far away distances if wrongly classified as main sequence, then they fall at closer distances with relatively high extinction values. This could lead to confusions on whether the increase in the extinction is caused due to such spurious values or is due to the presence of a cloud. We can overcome this difficulty by sub-divide the field containing the cloud into smaller fields. While the rise in the extinction due to the presence of a cloud should occur almost at the same distance in all  the fields, if the whole cloud is  located at the same distance, the wrongly classified stars in the sub-fields would show high extinction not at same but at random  distances. In case of cores that are having small  angular sizes, to have sufficient number of stars to infer their distances, we included fields containing additional cores that are located spatially closer and show similar radial velocities. Here we assume that the cores that are spatially closer and have similar velocities are located almost at similar distances.


\citet{1998ApJS..117..387K} have made a large scale survey in $^{13}$CO ($J=1-0$) of the Gemini and Auriga regions ($170^{\circ} < l \leq 196^{\circ}$ and $-10^{\circ}\leq b < 10^{\circ}$) with velocity coverages of $-30 <V_{LSR} < +30~km~s^{-1}$. Though they have not detected L1570 in their survey, they have detected a number of clouds that have $V_{LSR}$ relatively closer to that of $-0.5~km~s^{-1}$ of L1570 \citep{1988ApJS...68..257C}. We selected those clouds that are located below $l=0^{\circ}$ and have $V_{LSR}\leq\pm1~km~s^{-1}$. In Table \ref{tab:fields}, we present the field identification number, central galactic coordinates, cloud names and $V_{LSR}$ values as given by \citet{1998ApJS..117..387K} and the dark clouds associated with the regions. In Fig. \ref{fig:Av}, we identify three (F1-3) of the total six regions on $2.5^{\circ}\times1.5^{\circ}$ extinction map produced by \citet{2005PASJ...57S...1D}. Each field, F1-6, covered an area of $30^{\prime}\times30^{\prime}$. The $J$, $H$, and $K_{s}$ magnitudes of the stars were obtained from the 2MASS catalog \citep{2006AJ....131.1163S}. Only those stars that have photometric errors (which include the corrected band photometric uncertainty, nightly photometric zero point uncertainty, and flat-fielding residual errors) $\leq0.03$ magnitude and the photometric quality flag of ``AAA'' in all the three filters, i.e., signal-to-noise ratio (SNR) $>10$ were considered. The positions of L1570, L1576 and L1578 are identified and labeled. The contours are drawn at 1, 2, and 3 magnitude levels. The filled circles show the positions of stars classified as dwarfs used for estimating the distance.
 
\begin{figure}
\resizebox{9cm}{8cm}{\includegraphics{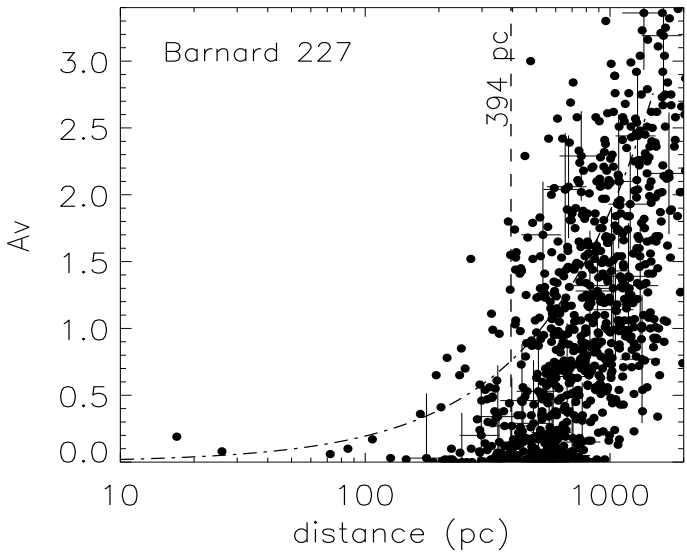}}
\caption{The $A_{V}$ vs. $d$ plot for all the stars obtained from the fields F1-F6 combined together towards L1570. The dashed vertical line is drawn at 394 pc inferred from the procedure described in \citet[][see the text for a brief description]{2010A&A...509A..44M}. The dash-dotted curve represents  the increase in the extinction towards the Galactic latitude of $b=-0.4591^{\circ}$ as a function of distance  produced from the expressions given by \citet{1980ApJS...44...73B}. The error bars are not shown on all the stars for better clarity.\label{fig:all}}
\end{figure}

\begin{figure}
\resizebox{9cm}{5.5cm}{\includegraphics{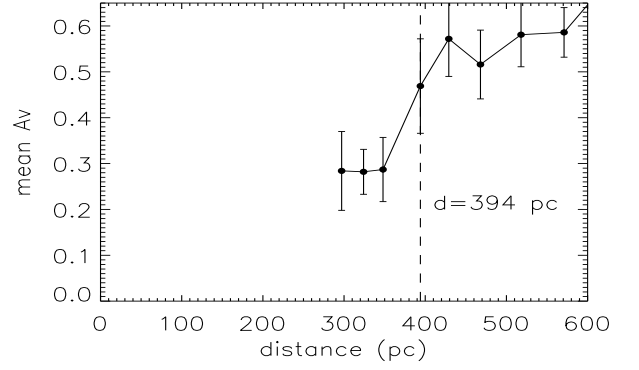}}
\caption{The mean values of $A_{V}$ vs. the mean values of distance plot for L1570 produced using the the procedure discussed in \citet[][see the text for a brief description]{2010A&A...509A..44M}.  The distance at which the first sharp increase in the mean value of extinction occur is taken as the distance to the cloud. The error bars on the mean $A_{V}$ values were calculated using the expression, $standard~deviation/\sqrt[]{N}$, where $N$ is the number of stars in each bin.\label{fig:hist}}
\end{figure}

In Fig. \ref{fig:all}, we present the $A_{V}$ vs. $d$ plot for all the stars obtained from the fields F1-F6 combined together towards L1570. The dash-dotted curve shows  the increase in the extinction towards the Galactic latitude of $b=-0.4591^{\circ}$ as a function of distance  produced from the expressions given by \citet{1980ApJS...44...73B}. The error bars are not drawn on all the stars for better clarity. A significant increase in the extinction is apparent at $\sim390$ pc. In order to determine distance to L1570, we first grouped the stars into distance bins of $bin~width = 0.18\times distance$. The centers of each bin are separated by the half of the bin width. Since there exist very few stars at smaller distances, the mean value of the distances and the $A_{V}$ of the stars in each bin were calculated by taking 1000 pc as the initial point and proceeded towards smaller distances. The mean distance of the stars in the bin at which a  significant drop in the mean of the extinction occurred was taken as the distance to the cloud and the average of the uncertainty in the distances of the stars in that bin was taken as the final uncertainty in distance determined by us for the cloud.  The vertical dashed line in $A_{V}$ vs. $d$ plots, used to mark the cloud distance,  is drawn at distance deduced from the above procedure. The error in the mean values of $A_{V}$ are calculated using the expression, $standard~deviation/\sqrt[]{N}$, where $N$ is the number of stars in each bin. From the above procedure, we determined a distance of $394\pm70$ pc to L1570 (see Fig. \ref{fig:hist}). 

\begin{figure}
\resizebox{9cm}{14cm}{\includegraphics{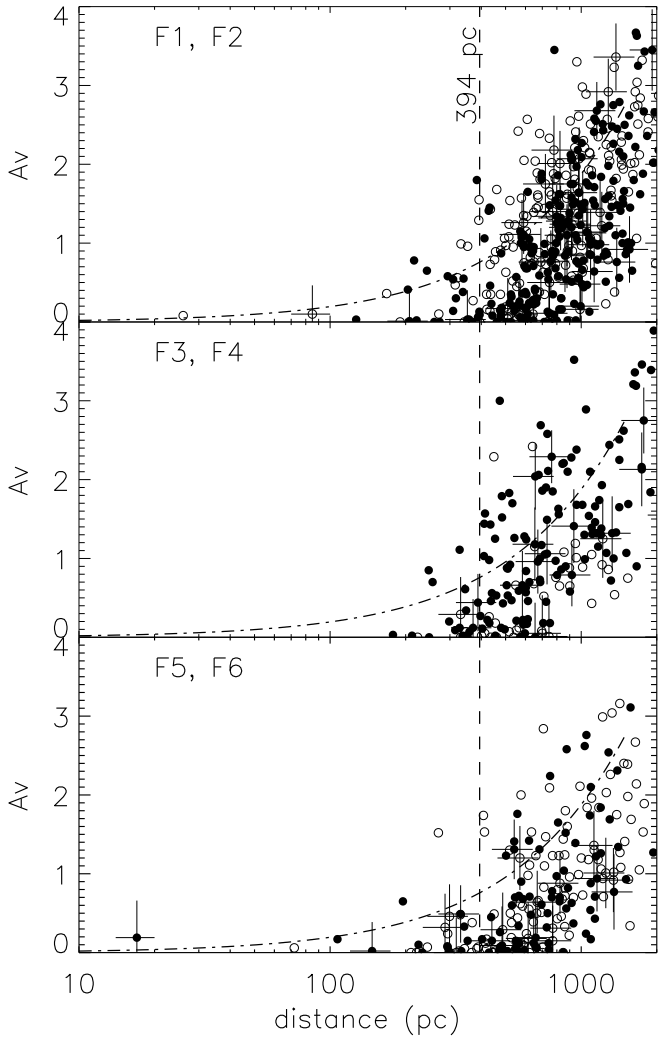}}
\caption{The $A_{V}$ vs. $d$ plots for the stars from the fields F1-F6 towards L1570. The stars from the fields F1, F3 and F5 are shown using filled circles and those from F2, F4 and F6 are shown using open circles. The dashed vertical line is drawn at 394 pc inferred from the procedure described in \citet{2010A&A...509A..44M}. 
The dash-dotted curve has the same meaning as in Fig. \ref{fig:all}.\label{fig:dist}}
\end{figure}

In Fig. \ref{fig:dist}, we show $A_{V}$ vs. $d$ plots for the stars from the  individual fields F1-F6. 
The dash-dotted curve has the same meaning as in Fig. \ref{fig:all}. The dashed vertical line is drawn at 394 pc. The stars from the fields F1, F3 and F5 are shown using filled circles and those from the fields F2, F4 and F6 are shown using open circles. In all the six fields, the increase in the extinction significantly above the values expected from the expressions of \citet{1980ApJS...44...73B} occurs consistently at or beyond $\sim390$ pc.  However, there exist evidence for the presence of a possible foreground dust layer at a distance of $\sim200$ pc. On the basis of 338 stars that are selected from six fields and classified as dwarfs, we estimated a distance of  $394\pm70$ pc to L1570.


\subsection{Dust properties using polarimetric and photometric data}\label{dust_prop}

\subsubsection{$P_{max}$, $\lambda_{max}$ and $R_{V}$ values}\label{pol_dust_properties}

Using our multi-wavelength data of the stars projected in the direction of L1570, we studied the properties of dust grains in the cloud. We obtained observations in $V(RI)_{C}$ filters for 42 stars and in $B,V(RI)_{C}$ filters for 15 stars. For our multi-wavelength observations we selected stars from various locations of the cloud so that the derived properties would represent the cloud as a whole.

We obtained $P_{max}$ and $\lambda_{max}$ using the weighted nonlinear least square fit to the measured polarization. We adopted $K$=1.15 for stars having data in $V(RI)_{C}$ pass-bands and $K$=1.66 $\lambda_{max}$ +0.01 ~\citep{1992ApJ...386..562W} for stars having data in $B,V(RI)_{C}$  pass-bands. We also computed the parameters $\sigma_{1}$\footnote{The values of $\sigma_{1}$ for each star are computed using the expression $\sigma_{1}^{2}=\sum(r_{\lambda}/\epsilon_{p\lambda})^{2}/(m-2)$; where $m$ is the number of colors and $r_{\lambda}=P_{\lambda} -­P_{max} \exp[-K~ln^{2} (\lambda_{max}/\lambda)$.}, the unit weight error of the fit for each star, which quantifies the departure of the data from the standard Serkowski law and $\overline{\epsilon}$, the dispersion of the polarization angle for each star normalized by the average of the polarization angle errors ~\citep[cf.][]{1993AJ....105..258M}. The Serkowski parameters namely, $P_{max}$, $\lambda_{max}$, $\sigma_{1}$ and $\overline\epsilon$ derived using $V(RI)_{C}$ \footnote{In order to check the consistency in the values of $P_{max}$ and $\lambda_{max}$ derived using four data points and three data points, we derived the values of $P_{max}$ and $\lambda_{max}$ for 15 stars that have $B,V(RI)_{C}$ data using $V(RI)_{C}$ only, i.e., leaving out the B band data. The results were then compared with those obtained using the data from all the four filters. The values obtained using the three and the four pass-band data are found to be consistent within the error.} wavelengths are given in columns 4, 5, 6 and 7 respectively, of Table \ref{vri_pol42} and those derived using $B,V(RI)_{C}$ are given in columns 4, 5, 6 and 7 respectively, of Table \ref{bvri_pol15}. 

\begin{figure}
\begin{small}
\vskip.1cm
\resizebox{8.5cm}{13cm}{\includegraphics{Sigma_vs_Pmax_and_Epsilon_vs_Lmax.epsi}}
\caption{ Upper panel: $\sigma_1$ vs. $P_{max}$ and Lower panel: $\overline\epsilon$ vs. $\lambda_{max}$. The stars represented in filled black circles are those with H$\alpha$ emission features. The star with $\sigma_{1}$ $\textgreater$ 1.5, $\epsilon \textgreater$ 4.0 and 0.4 $\textless \lambda_{max}$ $\textless$ 0.90  are identified and labeled with numbers from the Table\ref{r-band-127}. One star (\#23) with large values of $\sigma_{1}$ and $\overline\epsilon$ are indicated with arrow as those values are falling out of the plotted range.} 
\label{sig_vs_pmax_and_eps_vs_lmax}
\end{small}
\end{figure}

If the wavelength dependence of the polarization is well represented by the Serkowski law, $\sigma_{1}$ should not be greater than 1.5 because of the weighting scheme. A higher value ($\textgreater$ 1.5) could be indicative of intrinsic stellar polarization  \citep[][and the references therein]{1999AJ....117.2882W, 2008MNRAS.391..447F, 2011MNRAS.411.1418E, 2012MNRAS.419.2587E}. The rotation of polarization angle with wavelength ($\overline\epsilon$) also indicates the presence of an intrinsic polarization or a change of  $\lambda_{max}$ along the line of sight \citep{1974AJ.....79..565C,1974ApJ...187..461M}. Systematic variations with wavelength in the position angle of the interstellar linear polarization of star light may also be indicative of multiple dust layers with different magnetic field orientations along the line of sight \citep{1997ApJ...487..314M}. Following the above stated circumstances, we considered stars  that showed $\sigma_{1} \textgreater 1.5$ and $\overline\epsilon \textgreater 4.0$ (here, we considered only those stars that show $\overline\epsilon$ values falling away from the normal distribution followed by the rest of the stars)  as probable candidates to have either intrinsic polarization and/or rotation in the polarization angle. As shown in the upper and lower panels of the Fig \ref{sig_vs_pmax_and_eps_vs_lmax}, we considered ten stars (\#20, 23, 30, 47, 55, 56, 59, 83, 90, 97) as the candidates to have either intrinsic polarization and or rotation in their polarization angles. 

Another criterion to detect intrinsic stellar polarization is based on the value of $\lambda_{max}$. A star having $\lambda_{max}$ much lower than the average value of the ISM \citep*[0.545 $\mu$m;][]{1975ApJ...196..261S} is considered as a candidate to have an intrinsic component of polarization \citep*{1998AJ....116..266O}. In the present study we found only one star, \#23 (whose $\sigma_{1}$ =16.9 and $\overline\epsilon$=17.8), to show a much lower value of $\lambda_{max}$=0.33$\pm$0.02$\mu$m. We also consider two stars, \#24 and \#104, as peculiar because of their $\lambda_{max}$ being greater than 0.85$\mu$m. 


\begin{figure}
\vskip.1cm
\resizebox{8cm}{10cm}{\includegraphics{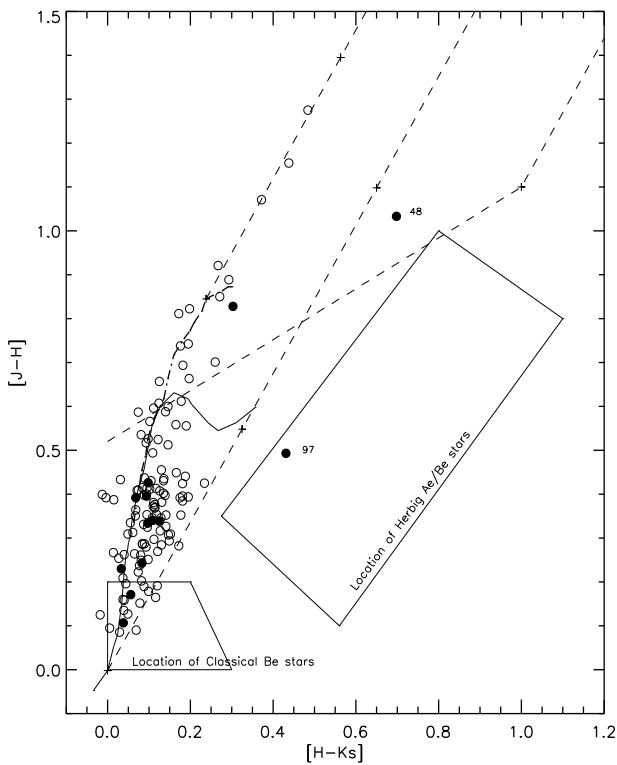}}
\caption{ $(J-H)$ vs. $(H-K)$ color-color diagram for all the observed 127 stars of L1570 with either single R-band and or with VRI or BVRI data sets. The data is taken from the ~\citep{2003yCat.2246....0C} catalog. The 2MASS data has been converted to California Institute of Technology (CIT) system using the relations provided by ~\citet{2001AJ....121.2851C}. The theoretical tracks for dwarfs (thin line) and giants (thick line) are drawn ~\citep*{1988PASP..100.1134B}. Reddening vectors (dashed lines) are also drawn ~\citep{1981ApJ...249..481C}. The location of Be stars ~\citep[cf.][]{1994A&A...290..609D}, and the location of Herbig Ae/Be stars ~\citep[cf.][]{2005AJ....129..856H} are also shown. The stars with $H\alpha$ emission in their spectra (labeled with their unique star id), the stars with probable intrinsic polarization and or rotation in their polarization angles and the stars with much smaller or larger $\lambda_{max}$ are shown with filled circles.
}
\label{NIRCCD}
\end{figure}

\begin{figure}
\vskip.1cm
\resizebox{9.0cm}{8.0cm}{\includegraphics{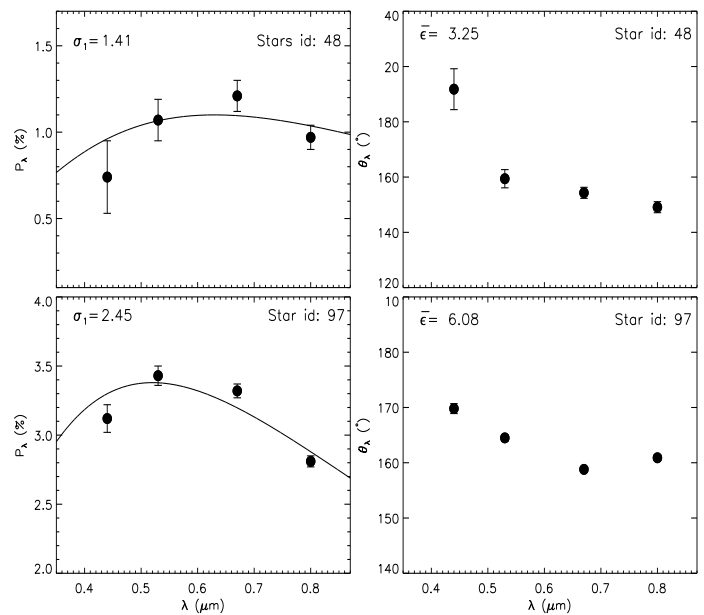}}
\caption{Left panels: Serkowski fit to the $\lambda$-dependent of polarization for the two stars \#48 and \#97. 
Right panels: $\lambda$-dependent of polarization angles for the same stars.}
\label{serk_48_97}
\end{figure}

The intrinsic polarization in a star could also be due to the asymmetric distribution of circumstellar material around the star in a disk or in a non-spherical envelop. The presence of circumstellar material around a star is inferred using NIR-CC diagram as shown in Fig \ref{NIRCCD}.  We constructed NIR-CC diagram using 2MASS JHKs magnitudes. The NIR excess sources occupy locations to the right of the reddening vector drawn for O-B spectral type stars. In Fig. \ref{NIRCCD}, open circles represent 127 stars that are observed by us. Only two stars namely, \#48 and \#97 (filled circles) show the clear presence of NIR excess. Interestingly, the stars \#48 and \#97 are  located at regions that are generally occupied by classical T-Tauri and Herbig Ae/Be stars respectively. In Fig \ref{serk_48_97} we show the variation of the degree of polarization and position angle for \#48 and \#97 as function of wavelength. As discussed earlier, both the stars show the presence of both intrinsic polarization and a rotation in their polarization angle. In both the cases, the polarization angle seems to decrease with the increasing wavelength which is similar to those observed by ~\citet*{1997ApJ...487..314M} towards the direction of Taurus dark cloud. 

\begin{figure}
\begin{small}
\vskip.1cm
\resizebox{9cm}{13.0cm}{\includegraphics{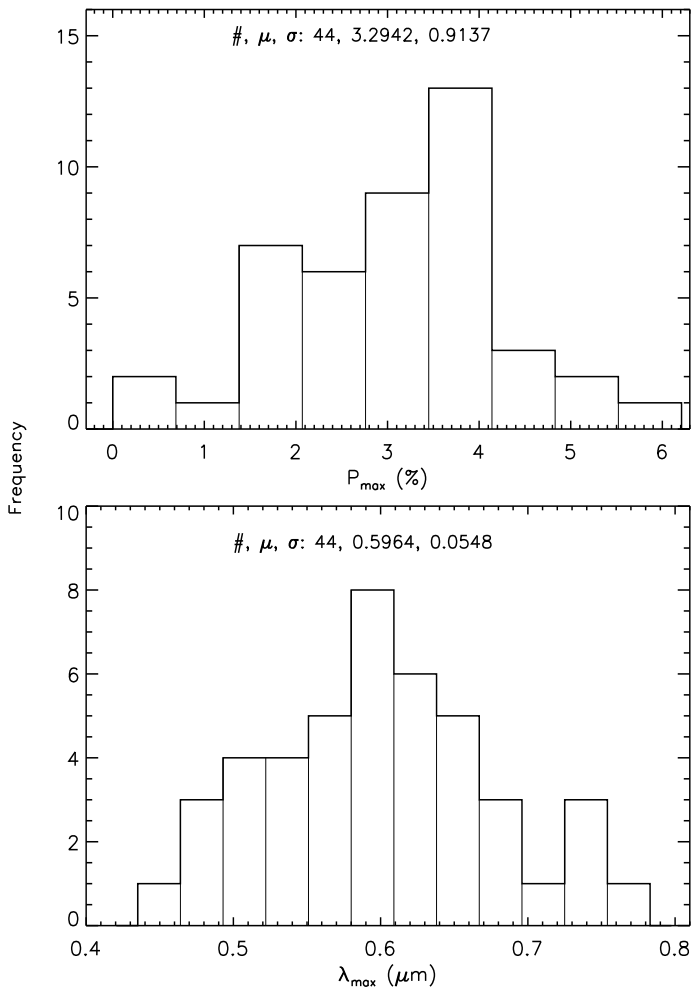}}
\caption{Distribution of $P_{max}$ (upper panel) and $\lambda_{max}$ (lower panel) for 44 stars. The mean and standard deviation values obtained using Gaussian fits to the data are also indicated.}
\label{pmax_lmax_gauss}
\end{small}
\end{figure}

Based on the values of $\sigma_{1}$, $\overline\epsilon$, i.e., with $\sigma_{1}$ $\textgreater$ 1.5 and/or $\epsilon$ $\textgreater$ 4.0 and/or lower or higher values of  $\lambda_{max}$ and/or the stars with possible NIR-excess based on NIR-CC diagram (Fig. \ref{NIRCCD}), we identified 13 stars (20, 23, 24, 30, 47, 48, 55, 56, 59, 83, 90, 97 and 104) that could possibly have intrinsic polarization and/or rotation in their polarization angles. These stars are excluded from our study of dust properties of L1570. Figure \ref{pmax_lmax_gauss} shows the frequency distribution of $P_{max}$ (upper panel) and $\lambda_{max}$ (lower panel) for 44 stars. The mean and standard deviation of $P_{max}$ and $\lambda_{max}$ are obtained by making Gaussian fits as 3.29$\pm$0.91 per cent and 0.60$\pm$0.05 $\mu$m respectively. Using the relation $R_V = (5.6 \pm 0.3) \times \lambda_{max}$ \citep{1978A&A....66...57W}, the value of $R_V$ is found to be 3.4$\pm$0.3, which is slightly higher than the average value ($R_V$ = 3.1) for the Milky Way. 

\begin{figure}
\begin{small}
\vskip.1cm
\resizebox{8cm}{11cm}{\includegraphics{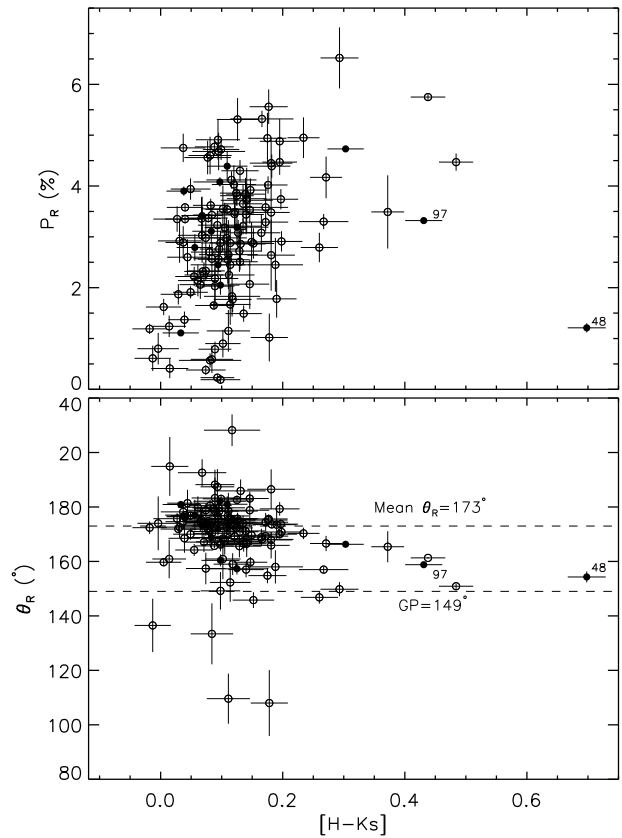}}
\caption{Upper panel: Degree of polarization in R-band vs. H-K$_{s}$ color. Lower panel: Polarization angle in R-band vs. H-K$_{s}$ color. Horizontal dotted lines denote the Galactic parallel (GP) and the mean value of the polarization angles at 149$\degr$ and 173$\degr$, respectively. 
The symbols are same as that of the Fig. \ref{NIRCCD}.}
\label{rp_rt_vs_hk} 
\end{small}
\end{figure}

In Figure \ref{rp_rt_vs_hk} we show the plot between the polarization and the polarization angle versus H-K$_{s}$ color for 114 stars. The thirteen stars with probable intrinsic polarization and or polarization angle rotation are shown with filled circles. As shown in the upper panel, the degree of polarization in R-band seems to increase with the H-K$_{s}$ color. This suggests that the main source of polarization is most likely the selective extinction by the aligned dust grains in the cloud. The lower panel suggests that the polarization angle for the majority of the stars is different from that of the Galactic parallel (149$\degr$) with a mean around 173$\degr$. However, there seems to be an indication of the position angles becoming aligned more with the Galactic parallel with increasing H-K$_{s}$ color. 
\begin{figure}
\begin{small}
\vskip.1cm
\resizebox{9cm}{10cm}{\includegraphics{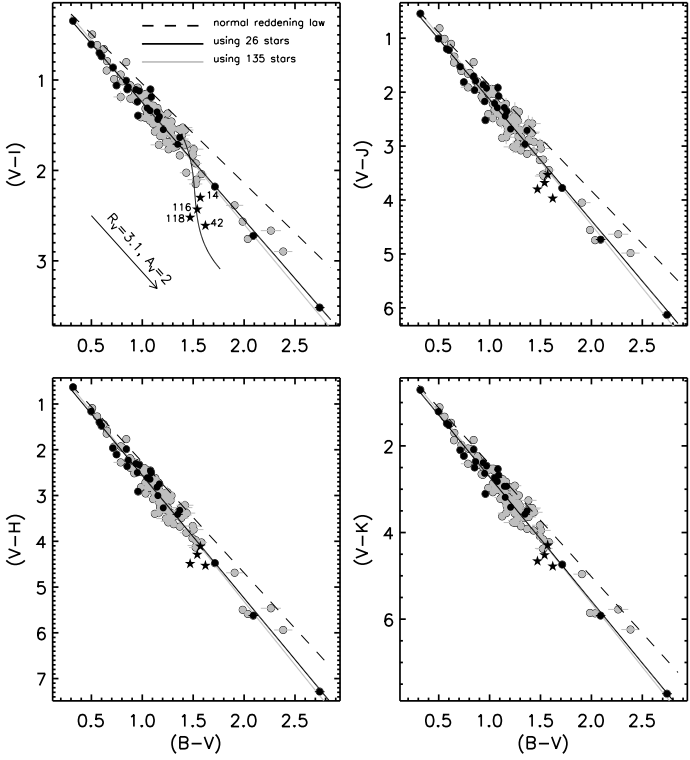}}
\caption{$(V-I)$, $(V-J)$, $(V-H)$, $(V-K)$ vs. $(B-V)$, two-color diagrams for the 135 stars (gray filled circles) having good photometric data (photometric errors $\textless$ 0.1 mag in BVRIJHK-bands) are plotted. The black filled circles denote 26 stars with both polarimetric and photometric data. The dashed lines are drawn using the color-color slopes mentioned in the Table \ref{slopes_TCDs} to represent the normal reddening law. Where as the thick and gray straight lines denote the error weighted straight line fits using 26 and 135  stars respectively. We excluded stars with possible intrinsic polarization (\#48, \#59, \#83, \#90 and \#97) from this analysis. We also excluded four stars, shown with filled star symbols, namely, \#14, \#42, \#116 and \#118 considering them as M-type dwarfs based on their location in $(V-I)$ vs. $(B-V)$ plot. The thick curve shows the  M dwarf locus taken from \citet{1998AJ....116..881P}. The reddening vector for a normal reddening law is drawn for $A_{V}$=2 mag.}
\label{TCDs}
\end{small}
\end{figure}

\subsection{$R_{V}$ value based on two-color diagrams} \label{phot_dust_properties}

The size of the dust grains can be constrained with the help of the parameter $R_{V}$. The mean $R_{V}$ for the Milky-way Galaxy is found to be 3.1. But observationally it is found that the $R_{V}$ is not fixed but rather vary from one line of sight to the other. For example towards the high latitude translucent molecular cloud direction of HD 210121 ~\citep*{1996ApJ...472..755L}, ~R$_{V}$ is found to be 2.1 where as towards the HD 36982 molecular cloud direction in the Orion nebula the R$_{V}$ is found to be 5.6. 

In order to investigate the nature of the extinction law in L1570, we used the two color-diagrams (TCDs) as described by \citet{2000PASJ...52..847P} and \citet{2003A&A...397..191P}  in the form of $(V-\lambda)$ versus $(B-V)$, where $\lambda$ is one of the wavelengths of the broad-band filters R, I, J, H, K or L, to separate the influence of the normal extinction produced by the diffuse ISM with average dust grain size from that of the abnormal extinction arising within regions having a peculiar distribution of dust sizes \citep[cf. ][]{1990A&A...227..213C, 2000PASJ...52..847P}. The $(V-\lambda)$ versus $(B-V)$ TCDs for the L1570 region are shown in Fig. \ref{TCDs}. We found 144 stars with both optical (BVRI) and 2MASS (JHK) data with good photometric quality (photometric errors $\textless$ 0.1 mag in $BVRIJHK$-bands). To characterize the dust grain size using TCD, we should use the stars that are reddened by the cloud material. In order to exclude any un-reddened M-type dwarfs from our sample (which may contaminate the reddened sample), we superposed the locus of the M-type dwarfs (continuous curve) obtained from \citet[][]{1998AJ....116..881P} in the $(V-I)$ vs. $(B-V)$ plot (top left panel). As shown in the top left panel, we identified four stars (\#14, \#42, \#116 and \#118 shown with filled star symbols) that are consistent with the intrinsic colors corresponding to the M-type dwarfs. However, the same un-reddened or reddened M-type dwarfs projected on the cloud could be used to determine a minimum distance to L1570 ~\citep[][]{1998AJ....116..881P}. The reddening\footnote{Reddening values are estimated by tracing back their $(B-V)$ and $(V-I)$ colors on to the intrinsic locus of M-type dwarfs by using the relations: $(B-V)_{0}$=$(B-V)-A_{V}$ $\times$ $R_{V}$; $(V-I)_{0}$= $(V-I)-$1.25$A_{V}$ $\times$ $R_{V}$, where $A_{V}$ is the reddening and $R_{V}$=3.1 (which is valid  for the diffuse ISM).} values for two stars \#14 and \#46 are found to be 0.20 and 0.25 mag, respectively. Where as the reddening values for \#116 and \#118 are assumed to be 0.0 mag. The spectral types of these four stars are found to be M2.5V, M3V, M3V and M3V. Using the absolute magnitudes, V-band magnitudes and the reddening values the distances are estimated as 68 pc, 239 pc, 190 pc and 169 pc for the stars \#14, \#42, \#116 and \#118, respectively. From this we conclude that L1570 is certainly located beyond $\sim240$ pc.

In addition to the five stars with possible intrinsic polarization (\#48, \#59, \#83, \#90 and \#97), the four M-type stars (\#14, \#42, \#116 and \#118; shown with filled star symbols) are also not used in the fit. Hence, out of 144 stars with photometric data, we used 135 stars (filled circles in gray) in TCD. Of these 135 stars, we found 26 stars with both polarimetric and photometric data (filled black circles). Error weighted straight line fits were performed for the two-color distributions by using 26 and 135 stars separately and the fitted slopes for each color-color distribution are mentioned in Table. \ref{slopes_TCDs}. In the fig. \ref{TCDs} the dotted line corresponds to the normal reddening law, where as the gray and thick lines correspond to the error weighted fitted slopes using the 26 and the 135 stars respectively. 

To derive the value of $R_V$ for the cloud region L1570, we use the approximate relation \citep[see][]{1981A&AS...45..451N},
\begin{equation}\label{eqnrv}
R_{cloud} = \frac{m_{cloud}}{m_{normal}} R_{normal}
\end{equation}
where $m_{cloud}$ and $m_{normal}$ are the slopes of the two-color combination for the stars towards the cloud region and for the MS stars in the normal diffuse ISM (taken from the stellar models by \citealt{1994A&AS..106..275B} and see also Table 3 of \citealt{2003A&A...397..191P}) respectively. $R_{normal}$ is taken as 3.1. Using the equation \ref{eqnrv} and the slopes (cf. Table \ref{slopes_TCDs}) for two cases with 26 and 135 stars, the weighted mean value of $R_{cloud}$ is estimated to be 3.53$\pm$0.02 and 3.64$\pm$0.01 respectively. Within the error both the sets of stars yield a similar value of $R_{V}$ thereby indicating the presence of a possible anomalous reddening law towards L1570 . The weighted mean of $\lambda_{max}$ (cf. section \ref{pol_dust_properties}) also indicated a slightly anomalous reddening law possibly due to the presence of bigger dust grains at the regions traced in our polarimetric and photometric observations. 

\subsection{Spectral Energy Distribution of two H$\alpha$ sources}\label{halpha_stars}

We produced spectral energy distribution (SED) for the stars \#48 and \#97 using BVRI (our photometric observations), JHKs ~\citep[2MASS;][]{2003yCat.2246....0C} and WISE ~\citep{2012yCat.2311....0C}  photometric data. The SEDs thus produced are shown in Fig. \ref{sed_48_97}.  A straight line (dash line) is fitted to the data between 2 to 10$\mu$m to find the spectral index ($\alpha$=-${\rm \frac{d log(\lambda F_{\lambda})} {d log(\lambda)} }$) and to check whether these stars could possess any excess in the mid-infrared region of their SEDs because of the circumstellar disks around them. The $\alpha$ values of the stars \#48 and \#97 are similar to those of a low mass PMS star (CTTs/Class II) and an intermediate PMS star (HAe/Be) as their $\alpha$ values are -1.01$\pm$0.16 and -1.82$\pm$0.12 respectively. The classification scheme is adopted from \citet{2006AJ....131.1574L}. 

Bluer parts of the SEDs were closely matched with the reddened Kurucz model spectra (dotted curves) corresponding to the spectral types of K4V (upper panel) and B4V (lower panel). The model spectra are reddened using the following relation: \citep{2007ApJ...663..320F}
\begin{equation}\label{eqn_modelspectra}
f_\lambda = F_{\lambda} (\theta_{R})^{2} 10^{-0.4 ~A_{\lambda}}
\end{equation}
where $F_{\lambda}$ is the intrinsic stellar surface flux obtained from Kurucz models{\footnote{The Kurucz stellar flux can be obtained from: http://www.stsci.edu/science/starburst/Kurucz.html}, $\theta_{R}$ $\equiv$ R/d is the stellar angular radius  (where R is the radius of the star and d is the distance) and $A_{\lambda}$ is the reddening at wavelength $\lambda$ which is represented using the mean $R_{V}$-dependent extinction law of the form $A(\lambda)/A_{V}=a(x)+b(x)/R_{V}$ \citep*{1989ApJ...345..245C}. We used the following relation to estimate the scaling factor \rm{$\gamma$} or $\theta_{R}^{2}$. The following relation actually uses the Kurucz model flux of the appropriate spectral type and scaled to the observed visual magnitude 
\citep[see][]{2004BASI...32..151S}
\begin{equation}\label{scaling_factor}
{\rm \gamma} \times K(5500\AA) \times e^{-\tau} = 3.46\times10^{-9} \times 10^{-0.4 V}
\end{equation}
where K(5500\AA) is the Kurucz flux at 5500\AA, V is the V-band magnitude and $\tau=A_{V}$/1.0863. The SEDs of two stars were fitted visually for a given combination of a spectral type (Kurucz model with temperature, gravity with solar metallicity), reddening $E(B-V)$, $R_{V}$ (hence $A_{\lambda}$) and V-band magnitude. While the star \#48 is found to be best fitted at $T_{eff}$=4500 K, g=4.5, $E(B-V)$=0.265 and $R_{V}$=3.1, the star \#97 is found to be best fitted at $T_{eff}$= 17000 K, g=4.0, $E(B-V)$=1.21 and $R_{V}$=3.46, corresponding to the spectral types of K4V and B4V respectively. This is consistent with the spectral types derived from their respective spectrum. 

Using $B$ and $V$ magnitudes (from our photometry) and spectral types of K4V and B4V, we estimated a distance of  $\sim400$ pc for \#48 ($B=17.33, V= 15.97, M_{V}=7.02, E(B-V)=1.36-1.07=0.29, A_{V}=0.29\times3.1$=0.9) and $\sim1$ kpc for \#97 ($B=14.75, V= 13.41, M_{V}=-1.52, E(B-V)=1.34+0.19=1.53, A_{V}=1.53\times3.1$=4.74). The star \#48 could probably be located just in front of L1570 since the extinction towards this star of 0.9 mag is consistent with value that is expected at that distance (see Fig. \ref{fig:all}) evaluated using the expression given by \citet[][]{1980ApJS...44...73B}. However, the exact nature of this star (whether a pre-main sequence or a normal emission type) is still unclear and requires further investigation. The star \#97 is clearly a background star which is supported by the presence of DIBs that are caused, most likely, by the material in the cloud. 

\begin{figure}
\centering
\resizebox{8.5cm}{8.5cm}{\includegraphics{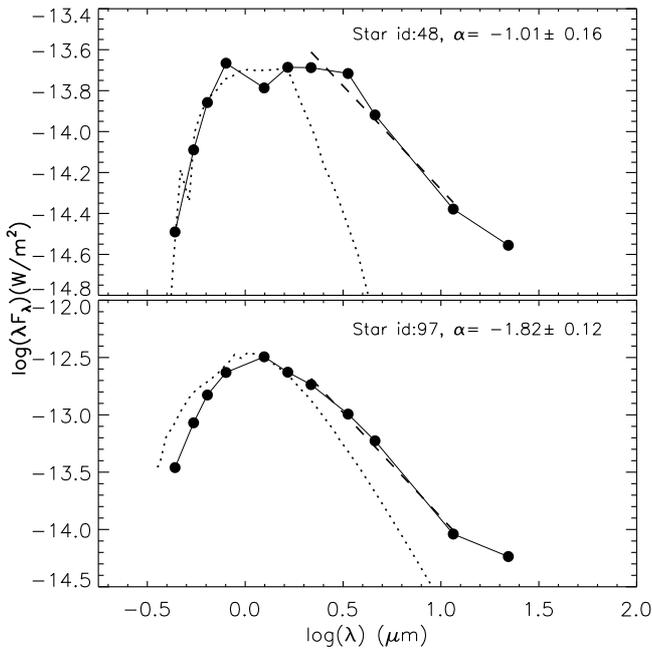}}
\caption{Spectral energy distribution ($log(\lambda F_{\lambda})$ versus $log(\lambda)$) of the stars \#48 and \#97 using non-simultaneous BVRI (see table \ref{photdata}), JHKs (2MASS, ~\citealt{2003yCat.2246....0C}) and WISE ~\citep{2012yCat.2311....0C}  photometric data sets (filled circles). A straight line (dash line) is fitted to the data between 2 to 10$\mu$m to derive the spectral index ($\alpha$=${\rm -\frac{d log(\lambda F_{\lambda})} {d log(\lambda)} }$).}\label{sed_48_97}
\end{figure}

\subsection{The polarization efficiency}\label{pol_effi}

The polarization efficiency of the dust grains towards a particular direction/line of sight is defined as the degree of polarization produced for a given amount of extinction. The efficiency of polarization produced depends on both the properties of the grains and the degree of alignment of these grains. Mie calculations for infinite cylindrical particles with dielectric optical properties that are perfectly aligned with their long axes parallel to each other and perpendicular to the line of sight place a theoretical upper limit on the polarization efficiency of the grains due to selective extinction. This upper limit is found to be $P/A_{V}\leq14 \%$mag$^{-1}$ \citep{2003dge..conf.....W}. The observational upper limit on $P/A_{V}$ is, however, found to be $\approx3 \%$mag$^{-1}$ \citep{2003dge..conf.....W} a factor of four less than the predicted value for the ideal scenario. 

\begin{figure}
\resizebox{8.725cm}{14cm}{\includegraphics{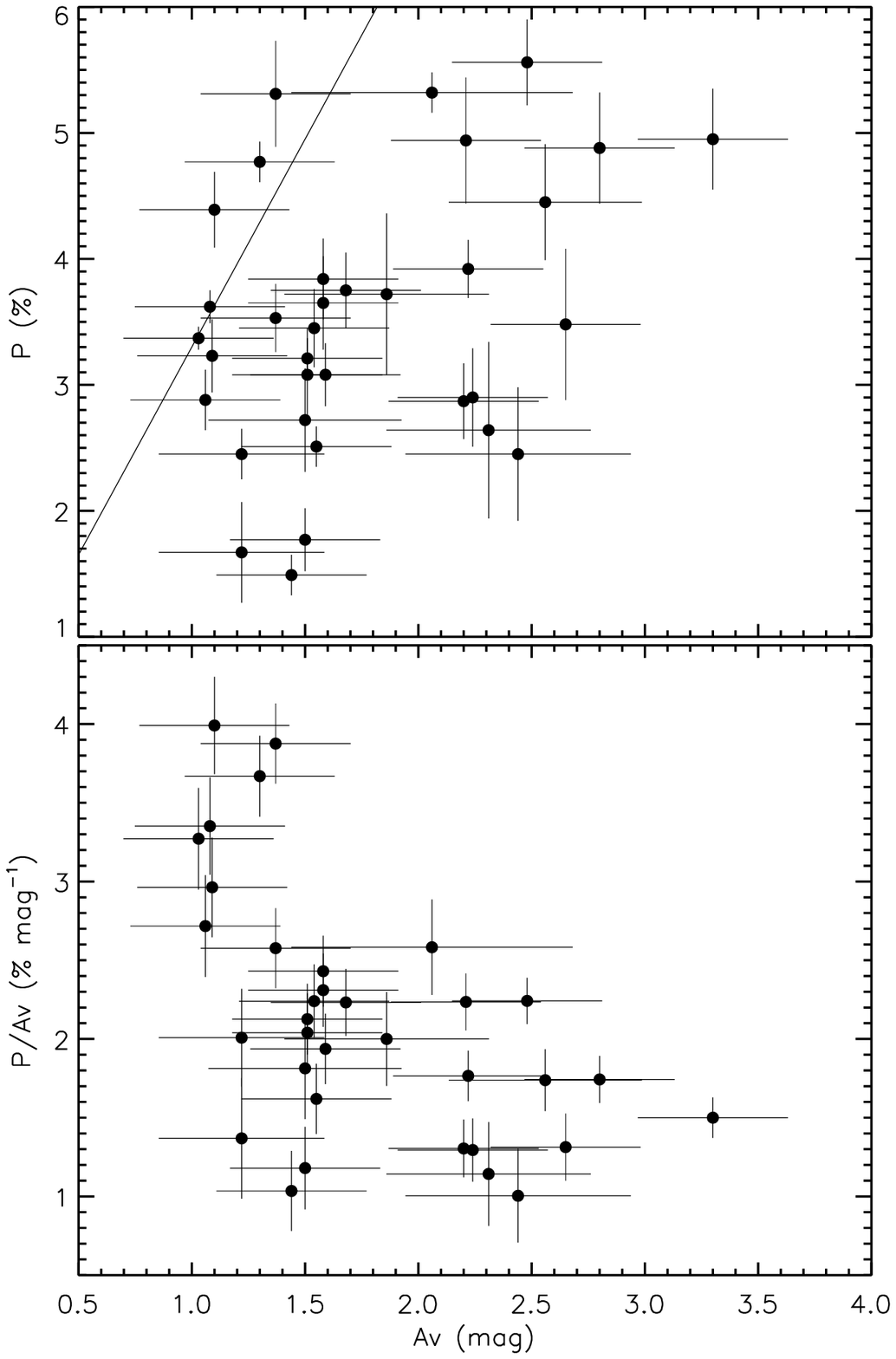}}
\caption{Upper panel shows the plot of degree of polarization (P) vs. total visual extinction (Av) derived from the method described in the section \ref{sub:dist}. We have selected only those sources for which Av/$\sigma$Av $\geq$ 3. Lower panel shows the polarization efficiency (P/Av) vs. visual extinction Av. The solid line represents the observational upper limit of $P/A_{V}=3$.\label{PAv}}
\end{figure}

Using the method outlined in the section \ref{sub:dist}, we could estimate $A_{V}$ for a total of 82 stars. In Fig. \ref{PAv} we present $P/A_{V}$ versus $A_{V}$ (upper panel) and P versus $A_{V}$ (lower panel) plots. We plotted only those sources for which Av/$\sigma_{Av}\geq$ 3. The plot in the upper panel shows that the polarization efficiency drops with increasing extinction. The plot of P versus $A_{V}$ shows that majority of the data points lie on or below the line representing the usual relation $P/A_{V}\approx3\%$ mag$^{-1}$ implying that  the characteristics of material composing L1570 is consistent with that of the diffuse ISM. The solid line in the lower panel of Fig. \ref{PAv} represents the observational upper limit of $P/A_{V}=3\%$mag$^{-1}$. 


\subsection{Magnetic field geometry of L1570}

Although the polarization of the stars located behind the cloud may be dominated by the dust associated with the cloud, the observed polarization will be a superposition of a component due to the dust located foreground to the cloud and another component due to the dust associated with the cloud. To evaluate the polarization produced only by the dust associated with the cloud, we need to subtract the foreground component from the observed polarization of the stars. This is essential to infer the true magnetic field orientation of the cloud and to study the magnetic field direction as a function of other cloud properties like the structure, outflow directions and the kinematics.

\subsubsection{Foreground dust polarization}\label{sec:foreground}

\begin{figure}
\centering
\resizebox{8.5cm}{12cm}{\includegraphics{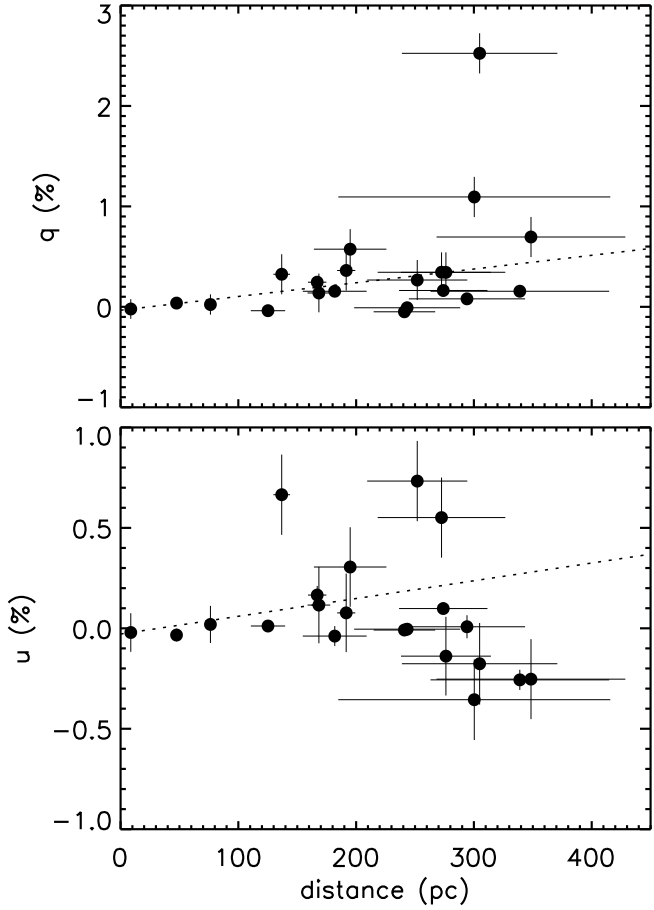}}
\caption{$q$ (upper panel) and $u$ (lower panel) versus distance plots using the stars distributed in a circular radius of 10$^\circ$ about L1570. Straight line fits were performed to estimate the foreground stokes parameters at the cloud possible minimum distance of 324 pc. 21 stars with available polarization measurements \citep{2000AJ....119..923H} and distance information \citep{2007A&A...474..653V} were only used. Care was taken to not to include stars that show
emission lines or are in binary systems as given in SIMBAD.\label{fig:uq_dist}}
\end{figure}

One way to evaluate the foreground dust component is to determine the behavior of polarization with distance up to the distance of the cloud. Even though, the distance and polarization information is known for 82 stars (Sec \ref{pol_effi}) (whose distance values are well beyond $\sim 400$ pc), the distance values for four M type stars (Sec \ref{phot_dust_properties}) are known (for which the polarization data is not available) and the distance of one H$\alpha$ emission star \#48 is $\sim 400$ pc (Sec \ref{halpha_stars}) (which exhibits NIR-excess (Fig \ref{NIRCCD}), intrinsic polarization and rotation in the polarization angles (Fig. \ref{serk_48_97})), none of the these stars can be used to estimate the foreground polarization. Therefore, we searched for stars within a circular region of radius $ 10^{\circ} $ about L1570 that have both polarization and distance measurements available in the literature. We obtained 21 stars that have polarization measurements in \citet{2000AJ....119..923H} catalog. Care was taken to not to include stars that show emission lines or are in binary systems as given in SIMBAD. The distance to stars are estimated using the Hipparcos parallax measurements available in the catalog by \citet{2007A&A...474..653V}. In Fig. \ref{fig:uq_dist}, we show the values of the Stoke's parameters $q$ and $u$ as a function of their distances. The uncertainty in our distance estimation of L1570 gives a minimum distance of 324 pc. Therefore we estimated the Stoke's parameters $q_{fg}$ and $u_{fg}$ values representing the foreground dust component at 324 pc by making a fit to the data points as shown in Fig. \ref{fig:uq_dist}. The estimated $q_{fg}$ and $u_{fg}$ at 324 pc are 0.3763 and 0.0625 respectively. These values were subtracted from the corresponding Stoke's parameters  of our 127 objects to get foreground corrected $q_{in}$ and $u_{in}$ values. From these, we calculated  the intrinsic degree of polarization and position angle for 127 objects. No significant change was noticed in the observed polarization results after making the correction for the foreground contribution.

\subsubsection{Magnetic field geometry}\label{mf_structure}

\begin{figure}
\centering
\resizebox{8.5cm}{8.5cm}{\includegraphics{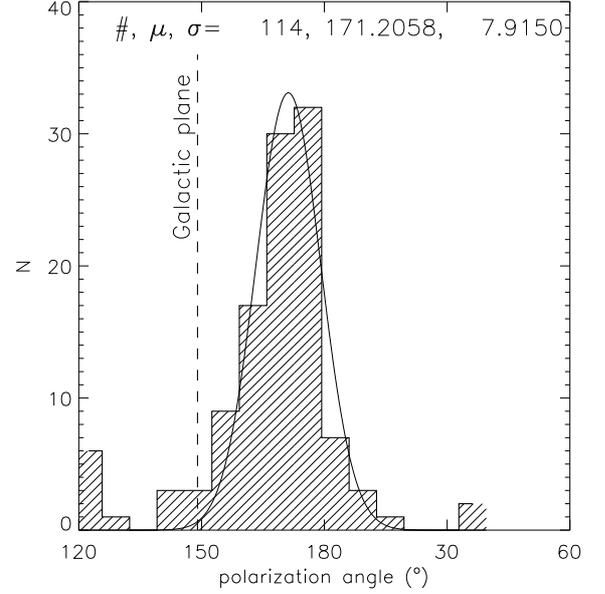}}
\caption{The frequency distribution of polarization angles of 114 stars of L1570 after the removal of foreground contribution. \label{fig:hist_ang_sub}}
\end{figure}
\begin{figure*}
\centering
\resizebox{14cm}{14cm}
{\includegraphics{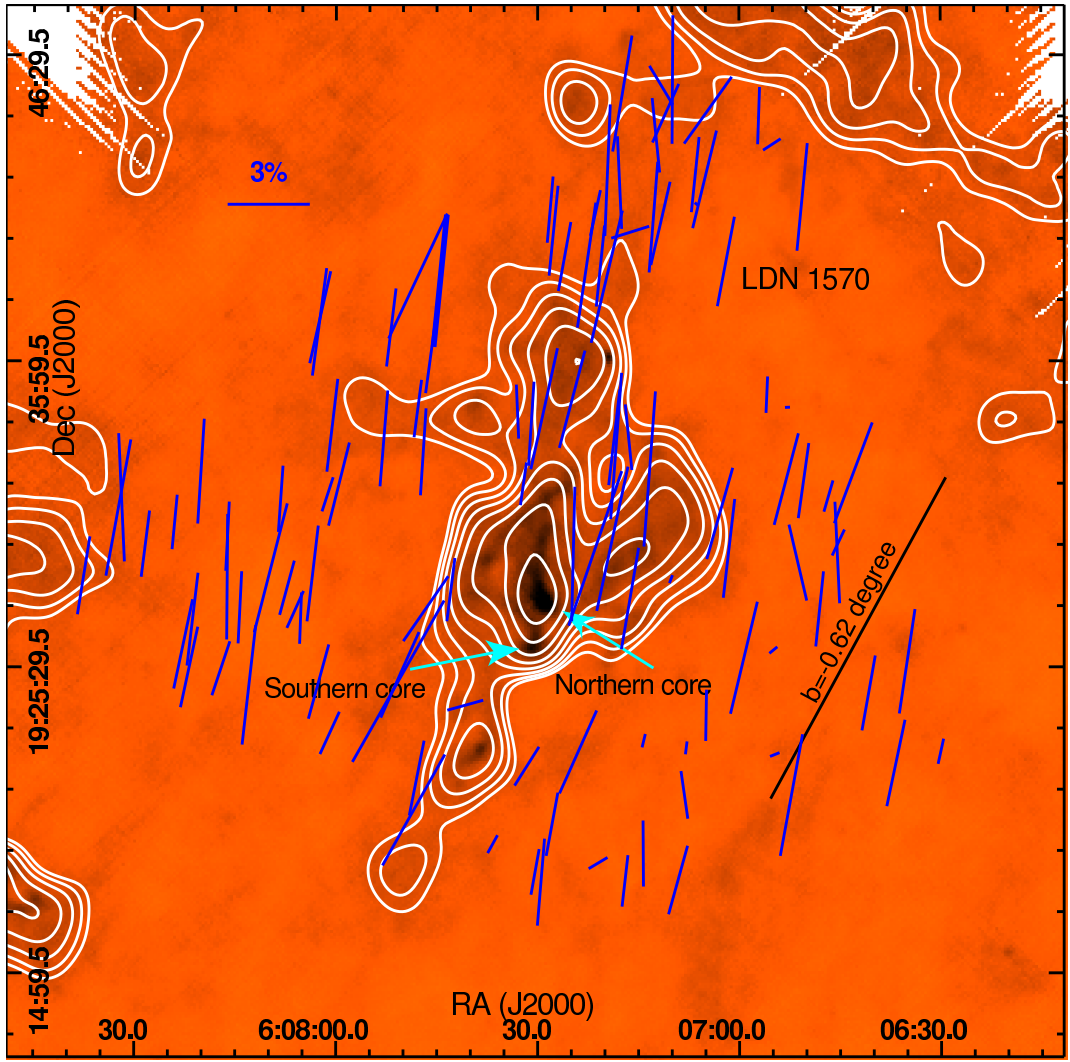}}
\caption{The polarization vectors (blue) of 127 stars (after removing foreground contribution) are over-plotted on the false color image of {\it HERSCHEL} 250 micron SPIRE \citep{2010A&A...518L...3G} Photometer Short (250$\mu$m) Wavelength Array (PSW) of the field containing L1570. The length of the vectors corresponds to the degree of polarization and the direction of the orientation corresponds to the polarization angle of the stars measured from the North increasing towards the East. A blue vector with a 3 percent polarization is drawn for a reference. The white contours represent the dust emission at 250$\mu$m at a minimum and maximum levels of 0.08 Jy/beam and 8.55 Jy/beam,  respectively. A color version of the figure is available in online journal. \label{fig:herschel_polvec}}
\end{figure*}

The mean value of the degree of polarization, after removing the foreground contribution as discussed in the previous section, is found to be 2.7\%  and a standard deviation of 1.2\%. Figure \ref{fig:hist_ang_sub} shows a peak in the distribution of polarization position angle centered at a mean of 171$^{\circ}$, with a standard deviation of 8$^{\circ}$ obtained by making a Gaussian fit. The thirteen stars with evidence for the presence of intrinsic component of polarization has been excluded from the analysis. The dispersion in the position angles is found to be significantly lower, similar to the dispersion of position angles observed towards the bowl region of the Pipe nebula which is suggested to be in a primordial evolutionary state \citep{2008A&A...486L..13A}.

In Fig. \ref{fig:herschel_polvec}, we present the polarization map superposed on the false color image of {\it HERSCHEL} 250 micron SPIRE \citep{2010A&A...518L...3G} Photometer Short (250$\mu$m) Wavelength Array (PSW) of the field containing L1570. The white contours represent the angular extent of the cloud in 250$\mu$m. The minimum and maximum contour levels correspond to 0.08 Jy/beam and 8.55 Jy/beam,  respectively. The contours indicate that the cloud is elongated along the north-south direction in a manner that the southern part of the cloud is oriented almost parallel to the Galactic parallel  ($\sim149\degr$ at $b=-0.62\degr$) where as the northern part seems to be slightly bent towards the Galactic plane ($b=0\degr$). The magnetic field geometry in L1570 also seems to follow the large scale structure seen in Fig. \ref{fig:herschel_polvec}. Towards the southern parts the field seems to be almost parallel to the Galactic parallel ($b=-0.62\degr$) and towards the northern parts the field lines are bent by $\approx20\degr$ towards the Galactic plane ($b=0\degr$) following the cloud structure. 

The morphology of L1570 in the two-dimensional map of the cloud in $^{13}$CO made by \citet{1985ApJ...297..436A} looks very much similar to the one shown in the Fig. \ref{fig:herschel_polvec} \citep[also see the 8 $\mu$m shadow image of L1570 produced by \textit{Spitzer} telescope; ][]{2009ApJ...707..137S}. The velocity structure of L1570, made by \citet{1986ApJ...303..356A}, gives a complex picture of the region. L1570 consists of sub-condensations towards the north and the south-southeast of the main body. Along the declination strip, the northern condensation is prominent and along the northwest-southeast strip, the southern condensations are prominent. Based on their results, \citet{1986ApJ...303..356A} suggested that the velocity structure of L1570 originates from the fragmented nature of the gas complex with relative velocity differences of 0.5 km s$^{-1}$ between the clumps.  Also towards a number of positions in L1570, they identified the presence of secondary spectral features within 2 km s$^{-1}$ of the L1570 main lines indicating further fragmentation at scales that were not resolved in their observations. In Fig. \ref{fig:herschel_polvec_smallreg}, we show the central region of L1570 produced by the HERSCHEL with polarization vectors over-plotted. The condensations identified initially by \citet{1986ApJ...303..356A} are clearly visible in Fig. \ref{fig:herschel_polvec_smallreg}. Both the northern and the southern condensations show filamentary structures. The filamentary structure seen towards the northern condensation seems to be aligned with the magnetic field lines.  

The magnetic field geometry of a molecular cloud is mainly governed by the relative dynamical importance of magnetic forces to gravity, turbulence and thermal pressure. If the magnetic fields are dynamically important, i.e., the support to the molecular cloud against gravity is provided predominantly by the magnetic field, then the field lines would be aligned smoothly and that the mean field orientation would be perpendicular to the major axis of the cloud \citep{1978prpl.conf..209M}. This is expected as the cloud tends to contract more in the direction parallel to the magnetic field than in the direction perpendicular to it. On the other hand, if the magnetic fields are dynamically unimportant compared to turbulence, then the random motions dominate the structural dynamics of the clouds, and the field lines would be dragged around by turbulent eddies \citep{1999ApJ...515..286B}. In such a case, magnetic fields would be chaotic with no preferred direction. 

In L1570, the relatively low dispersion in position angles implies that the magnetic field lines are aligned smoothly (see Fig. \ref{fig:hist_ang_sub}). The position angle of the major axis of the cloud is $\sim165\degr$ \citep[inferred from the optical DSS images; ][]{1988ApJS...68..257C}. Thus, the magnetic field lines in L1570 are oriented roughly parallel to the major axis contrary to what is expected for a magnetic field dominated scenario. \citet{1999ApJ...527..285B} and \citet{2001ApJ...562..852H} proposed that the molecular clouds are formed in large-scale, converging flows of diffuse atomic gas. Numerical simulations have shown that in a flow-driven cloud formation scenario, magnetic fields are believed to guide the flows to assemble the clouds whether through 
Parker instability \citep{1966ApJ...145..811P,1967ApJ...149..517P,1974A&A....33...73M,2009MNRAS.397...14M}, 
through turbulence in ISM, or during the swept-up of gas in spiral shocks \citep{2006ApJ...649L..13K, 2008MNRAS.383..497D}. 
\citet{2001ApJ...562..852H}, based on the models of \citet{1995ApJ...455..536P}, suggested that the field orientation with respect to the flows select the sites of cloud formation. The clouds will only form if the magnetic fields are aligned parallel to the flows. L1570 could have formed due to the converging flow of material along the field lines. This argument is supported by the orientation of filamentary structure along the field lines. The complex velocity structure in L1570 could be due to the collision of material in  converging flows from opposite sides along north-south direction guided by the magnetic field lines. Inference of velocity gradient in L1570 could confirm our argument. The magnetic field geometry of the high density regions of L1570, where we see substructures, in near infrared or submillimeter wavelengths would be highly desirable. 

\begin{figure}
\centering
\resizebox{8.275cm}{8.275cm}
{\includegraphics{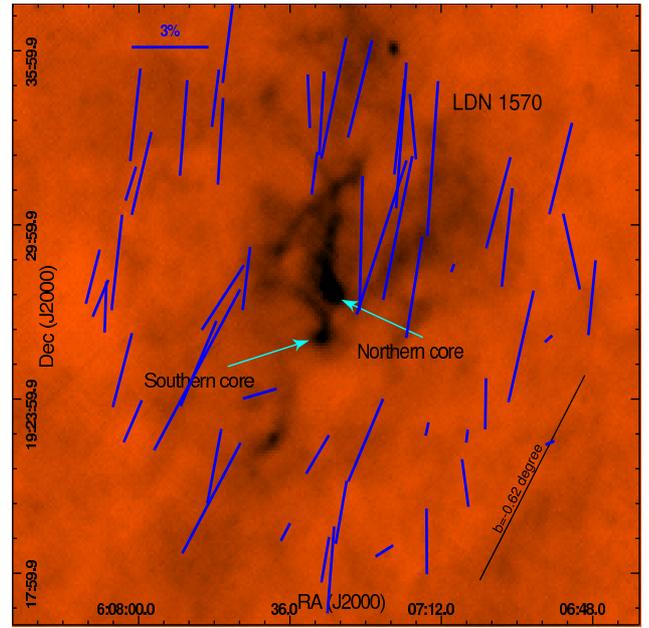}}
\caption{The polarization vectors (blue) of 54 stars that are distributed in a 10 arcmin radius around the cloud center are over-plotted on false color image of {\it HERSCHEL} 250 micron SPIRE \citep{2010A&A...518L...3G} Photometer Short (250$\mu$m) Wavelength Array (PSW) of the field containing L1570. The length of the vectors corresponds to the degree of polarization and the direction of the orientation corresponds to the polarization angle of the stars measured from the North increasing towards the East. A blue vector with a 3 percent polarization is drawn for a reference. A color version of the figure is available in online journal. \label{fig:herschel_polvec_smallreg}}
\end{figure}

\subsubsection{Magnetic field strength and Cloud stability}
To understand the cloud stability rendered by the magnetic fields against gravity and turbulence and 
to understand the importance of magnetic fields in cloud as well as star formation processes 
it is crucial to estimate the magnetic field strength \citep{2005ASPC..343..166H}.  

Magnetic field strength can be derived by estimating the turbulent dispersion, spectral line-widths and the density. 
Using the dispersion in the polarization position angles in the modified Chandrasekhar-Fermi (CF) formula \citep{1953ApJ...118..116C, 2001ApJ...546..980O}, we estimated the plane of the sky magnetic field strength. The value of  $n$(H$_{2}$) $\simeq598$ cm$^{-3}$ at 3 arc min distance (the distance at which the optical polarimetry is relevant) from the center of the cloud is adopted from \citet{1985ApJ...297..436A} and the line width value equal to 1.75 km~s$^{-1}$ found for $CO$ towards the core \citep{1991ApJS...75..877C}. Assuming these values, we found that the magnetic field strength in L1570, in the plane of the sky, is $\simeq$80 $\mu$G. 

However, CF method assumes that the dispersion in the polarization angles is caused purely due to the 
hydrodynamic turbulence. This is not entirely true as other non-turbulent components such as cloud collapse, 
differential rotation, gravitational collapse, expanding H{\sc ii} regions and, in addition, 
a component due to measurement error could also introduce the 
dispersion in the the magnetic fields as suggested by \citet{2009ASPC..417..257H}, \citet{2009ApJ...696..567H}, and \citet{2009RMxAC..36..137H}. 
And also, it is hard to separate the turbulent dispersion 
from that due to large scale structure (MHD waves along the spirals) and due to measurement errors. 

Field strength can be estimated using the method (analogous to that used by CF) 
proposed by \citet{2009ApJ...696..567H}. This method essentially determines 
the angular dispersion due to turbulence in molecular clouds, where the turbulent dispersion is distinguishable 
from dispersion due to the large-scale structure or the apparent dispersion due to measurement error. 
To separate the turbulent from non-turbulent components, 
we plotted $\langle\Delta{\theta^{2}}(l)\rangle^\frac{1}{2}$ 
\citep[][and references therein]{2008ApJ...679..537F,2010ApJ...723..146F,2010ApJ...716..893P,2012ApJ...751..138S}, 
the square root of the second-order structure function or 
angular dispersion function (ADF)\footnote{Angular dispersion function (ADF) is defined as the square root to 
the average of the squared difference between the polarization 
angles measured for all pairs of points ($N(l)$) separated by a distance $l$.}, as function of distance ($l$) 
as shown in the Fig. \ref{fig:strfunct_plot}. We have used the polarization angles of 127 stars to compute 
$\langle\Delta{\theta^{2}}(l)\rangle^\frac{1}{2}$, which 
gives information on the behavior of the dispersion of the polarization 
angles as a function of the length scale in L1570. 
Recently, it has been used as a powerful statistical tool to infer 
information on the relationship between the large-scale and the turbulent components of the 
magnetic field in molecular clouds \citep[see][and references therein]{2009ApJ...696..567H,2010ApJ...723..146F,2012ApJ...751..138S}. 

\begin{figure}
\centering
\resizebox{8.775cm}{6.00cm}
{\includegraphics{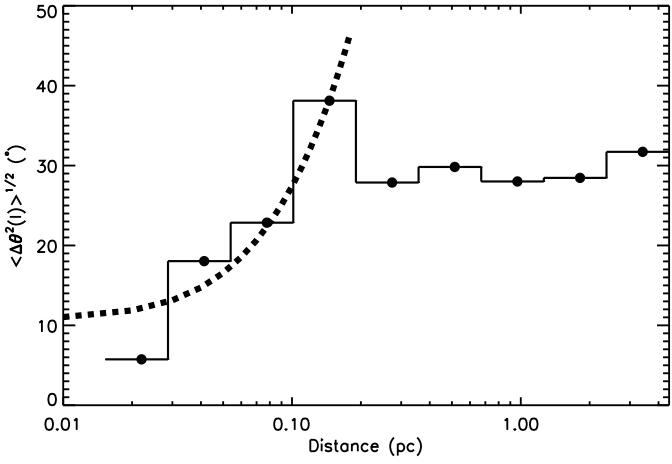}}
\caption{Square root of the second-order structure function (or angular dispersion function (ADF)), 
$\langle\Delta{\theta^{2}}(l)\rangle^\frac{1}{2}$ (degree),  
of the polarization angles versus distance ($l$) (pc) using 127 stars of L1570. 
The filled circles are the ADF values in each bin. The error bars are comparable to the 
size of the symbols. The turbulent contribution to the total angular dispersion function is determined 
by the zero intercept of the fit to the data at $l$=0. The measurement errors were removed before 
fitting the function to the data. Thick dotted line denotes the best fit to the data for distance less than 0.2 pc. 
\label{fig:strfunct_plot}}
\end{figure}

The square of the dispersion function can be approximated as follows \citep{2009ApJ...696..567H} 
\begin{equation}\label{ADF_function}
\langle\Delta{\theta^{2}}(l)\rangle_{tot}-{\sigma^{2}_{M}}(l)=b^{2} + m^{2}l^{2}, 
\end{equation}
where $\langle\Delta{\theta^{2}}(l)\rangle_{tot}$ is the dispersion function computed from the data. The 
quantity ${\sigma_{M}}^{2}(l)$ is the measurement uncertainties which is simply the average of the variances on $\Delta{\theta(l)}$ in each bin. The quantity $b^2$ is the intercept of a straight line fit to the data 
(after subtracting ${\sigma_{M}}^{2}(l)$). \citet{2009ApJ...696..567H} have derived the equation for $b^2$ 
to find the ratio of turbulent to the large-scale magnetic field strength:

\begin{equation}\label{turb_large_b}
\frac{\langle B^{2}_{t}\rangle^{1/2}}{B_{o}}=\frac{b}{\sqrt{2-b^2}}
\end{equation}

In Fig. \ref{fig:strfunct_plot}, we showed the ADF versus distance for L1570 region. The errors in each bin are 
comparable to the size of the symbols. 
Each bin denote $\sqrt{\langle\Delta{\theta^{2}}(l)\rangle_{tot}-{\sigma_{M}}^{2}(l)}$ i.e., 
the ADF corrected for the measurement uncertainties. Bin widths are 
in logarithmic scale. Only four points have been used in the linear fit, to make sure that 
the length scale ($l$) used in the fit (0.015 to 0.20 pc) is greater than the turbulent 
length scale ($\delta$) (which is of the order of 1 mpc or 0.001 pc; cf. \citealt{Lazarianetal2004}, \citealt{LiHoude2008}) 
and much less than the cloud length scale ($d \sim 1$ pc) i.e., $\delta~\textless~l~\ll~d$, 
to the data (eq. \ref{ADF_function}) versus distance squared.  
The fitted function 
is denoted with a thick dotted line. 
Since, our optical polarimetric observations have low resolution due to the 
available limited number of point sources, 
the minimum length we probed is $\simeq$ 15 mpc. 
The turbulent contribution to the total ADF is determined by the zero intercept of the fit to the 
data at $l$=0. The net turbulent component, $b$, is estimated to be $11\degr\pm3\degr$ 
(or 0.19$\pm$0.06 rad). The ratio of the turbulent to large-scale magnetic field strength 
($\sigma(\theta)=\langle B^{2}_{t}\rangle^{1/2}/{B_{o}}$) is computed using eq. \ref{turb_large_b} as 0.13$\pm$0.04. 
This suggest that the turbulent component of the field is very small compared to the 
non-turbulent component i.e., $B_{t} \ll B_{0}$. Based on this assumption \citet{2009ApJ...696..567H} show 
that the uniform/non-turbulent component of the field can be approximated by the following relation: 
\begin{equation}
B_{0}\simeq \sqrt{8\pi\rho} \frac{\sigma_{v}}{b}. 
\end{equation}

As already mentioned above, we have used $n(H_{2}$)$\simeq$598 cm$^{-3}$, 
velocity dispersion ($\sigma_{v}$)=0.74 $km s^{-1}$, and $b=11\degr\pm3\degr$. The density can be estimated as $\rho=n(H_{2})m_{H}\mu_{H_{2}}$, where $n(H_{2})$=hydrogen column density, 
$m_{H}$= is the mass of the hydrogen atom, and $\mu_{H_{2}} \approx 2.8$ is 
the mean molecular weight per hydrogen molecule. Using these values, $B_{0}$ is estimated 
to be $\simeq$90 $\mu$G. 
\citet[][cf. their equation 7]{2011ApJ...741...21C} have used the following relation to 
for estimating the dimension less magnetic critical index in terms of total extinction $A_{V}$ (mag) and $B_{0}$ ($\mu$G)  
\begin{equation}
\mu=2.4~A_{V}/B_{0}.  
\end{equation}
Using the mean extinction for the L1570 region as $A_{V}$=1.77 mag\footnote{We have used only 34 stars, whose 
$A_{V}/\sigma_{A_{V}} \geq 3.0$, to estimate the mean $A_{V}$. See section \ref{pol_effi} for more details.} 
and $B_{0}\simeq 90 \mu$G, we have 
estimated $\mu$ as 0.047. If $\mu\textgreater 1$ then the cloud is assumed to be at super critical. If $\mu\textless 1$ then the cloud is said to be under sub-critical. The estimated value of $\mu$ for L1570 indicates that the 
cloud is magnetically subcritical, hence we infer that the cloud is supported by the magnetic fields.
The value $\mu$ is also estimated using field strength $B_{0}\simeq 80 \mu$G (which is estimated using CF method) as 0.053. Thus, both the Hildebrand et al. and CF methods draw the similar conclusion that the cloud is under sub-critical. 


\section{Conclusions}\label{conclude} 
In this paper, we presented the results on dust properties and the magnetic field geometry towards a dark globule L1570 using multi-wavelength polarimetric and photometric observations. The following are the main conclusions of our present study.

\begin{itemize}
\item We estimated a distance of 394$\pm$70 pc to the cloud using near-IR photometry from 2MASS.

\item Based on our multi-wavelength polarimetric observations of 42 stars, the Serkowski parameters such as $P_{max}$, $\lambda_{max}$ and $\sigma_{1}$ and a polarization angle rotation indicator, $\overline\epsilon$, were determined. Stars with possible intrinsic polarization and (or) rotation in their position angles and (or) the stars with NIR-excess (based on their position in near-IR color-color diagram) were identified. These stars were excluded from our study of the dust properties and the magnetic field geometry. 

\item The mean values of $P_{max}$ and $\lambda_{max}$ are found to be 3.29$\pm$0.91 per cent and 0.60$\pm$0.05 $\mu$m respectively, slightly higher than the value, 0.545$\mu$m, corresponding to the general ISM. The value of $R_{V}$ estimated using the $\lambda_{max}$ is found to be $3.4\pm0.3$. Using $(V-I)$, $(V-J)$, $(V-H)$, $(V-K)$ vs. $(B-V)$ two-colour diagrams for the 135 stars, we evaluated the value of $R_{V}$ as 3.64$\pm$0.01. The $R_{V}$ values derived from both these methods show that the grain size in L1570 is slightly bigger than those found in the diffuse ISM.

\item We confirmed the presence of H$\alpha$ emission features based on our spectroscopic observations of two stars namely 2MASS J06071585+1930001 (\# 48) and 2MASS J06075075+1934177 (\# 97). These two stars were found to show NIR-excess also. The star \# 97 show some of the prominent diffuse interstellar bands in the  spectrum. The spectral types of these stars are found to be K4Ve and B4Ve respectively. We confirmed the spectral type determination of these sources by produced spectral energy distribution  using optical, near and far-IR data.

\item The magnetic field geometry of L1570 seems to follow the large scale structure seen in the 250$\mu$m image produced by the Hershel. Towards the southern parts the field seems to be almost parallel to the Galactic parallel ($b=-0.62\degr$) whereas towards the northern parts the field lines are bend by $\approx20\degr$ towards the Galactic plane ($b=0\degr$). The filamentary structure seen towards the northern condensation are found to be aligned with the magnetic field lines. Based on the morphology of the magnetic field lines with respect to the cloud structure, we believe that L1570 could have formed due to the converging flow of material along the field lines. Using the dispersion in the polarization position angles in the modified Chandrasekhar-Fermi formula, we estimated the plane-of-the-sky magnetic field strength towards the outer parts of L1570 as $\simeq 80\mu$G. Similarly, structure function analysis yield the magnetic field strength as $\simeq 90 \mu$G.  
 
\item Structure function analysis suggests that the large scale magnetic fields are stronger when compared with 
the turbulent component of magnetic fields in L1570 cloud region. The estimated magnetic field strengths using Hildebrand 
et al. and Chandrasekhar Fermi methods suggest that the L1570 cloud region is under sub-critical and hence could be 
strongly supported by the magnetic field lines.
  
\end{itemize}

\section*{Acknowledgments}
Authors are highly thankful to the referee Dr G. A. P. Franco for his insightful comments and suggestions which 
helped in considerable improvement of the manuscript. This publication makes use of data products from the Two Micron All Sky Survey, which is a joint project of the University of Massachusetts and the Infrared Processing and Analysis Center/California Institute of Technology, funded by the National Aeronautics and Space Administration and the National Science Foundation. We also used the images from the Digitized Sky Surveys which were produced at the Space Telescope Science Institute under U.S. Government grant NAG W-2166. The images of these surveys are based on photographic data obtained using the Oschin Schmidt Telescope on Palomar Mountain and the UK Schmidt Telescope. We acknowledge the use of NASA's \textit{SkyView} facility (http://skyview.gsfc.nasa.gov) located at NASA Goddard Space Flight Center.This research has also made use of the SIMBAD database, operated at CDS, Strasbourg, France. 
\bibliographystyle{aa}
\bibliography{myref}

\onecolumn 
\small
\begin{landscape} 
\begin{center}
\begin{longtable}{ccccccccc}
\caption[]{Observed polarized standard stars from S92} \label{result_standards} \\\hline 

\textbf{Star Name} & \textbf{$P_{B}\pm\epsilon$} & \textbf{$\theta_{B}\pm\epsilon$} & \textbf{$P_{V}\pm\epsilon$} &  \textbf{$\theta_{V}\pm\epsilon$} & \textbf{$P_{Rc}\pm\epsilon$} &  \textbf{$\theta_{Rc}\pm\epsilon$}  & \textbf{$P_{Ic}\pm\epsilon $} & \textbf{$\theta_{Ic}\pm\epsilon$}\\ 
\endfirsthead
\multicolumn{9}{c}%
{\tablename\ \thetable\ -- \textit{Continued from previous page}} \\
\hline

\textbf{Star Identification} & \textbf{$P_{B}\pm\epsilon$} & \textbf{$\theta_{B}\pm\epsilon$} & \textbf{$P_{V}\pm\epsilon$} &  \textbf{$\theta_{V}\pm\epsilon$} & \textbf{$P_{Rc}\pm\epsilon$} &  \textbf{$\theta_{Rc}\pm\epsilon$}  & \textbf{$P_{Ic}\pm\epsilon $} & \textbf{$\theta_{Ic}\pm\epsilon$}\\
\hline
\endhead
\hline \multicolumn{9}{r}{\textit{Continued on next page}} \\
\endfoot
\hline
\endlastfoot

(data of observation)      &  (per cent) & ($\degr$) &  (per cent) & ($\degr$) &  (per cent) & ($\degr$) &  (per cent) & ($\degr$)  \\ 
 (1)   &  (2)     &   (3)    &   (4)  &  (5)   &   (6)   &  (7)      &  (8)        &   (9)   \\ \hline
  \multicolumn{9}{c}{Standard stars observed using AIMPOL}\\
\hline
HD19820 (2009 November 23)  &   4.49  $\pm$   0.11  &    114.9  $\pm$      0.7 &   4.89  $\pm$   0.09  &    114.2  $\pm$  0.5 &  4.49  $\pm$   0.09  &    115.5  $\pm$      0.6 &   4.06  $\pm$   0.16  &    115.3  $\pm$      1.0  \\  
HD19820 (2009 November 24)  &   4.61  $\pm$   0.11  &    114.7  $\pm$      0.7 &   4.79  $\pm$   0.08  &    115.1  $\pm$  0.5 &  4.51  $\pm$   0.07  &    114.6  $\pm$      0.4 &   3.97  $\pm$   0.09  &    115.7  $\pm$      0.6  \\
HD19820 (2009 November 25)  &   4.72  $\pm$   0.11  &    115.6  $\pm$      0.7 &   4.94  $\pm$   0.08  &    114.9  $\pm$  0.4 &  4.61  $\pm$   0.07  &    114.4  $\pm$      0.4 &   3.86  $\pm$   0.10  &    113.4  $\pm$      0.7  \\ 
HD19820 (2009 November 26)  &   4.82  $\pm$   0.12  &    114.9  $\pm$      0.7 &   4.42  $\pm$   0.10  &    115.1  $\pm$  0.6 &  4.53  $\pm$   0.09  &    114.4  $\pm$      0.6 &   4.02  $\pm$   0.08  &    114.1  $\pm$      0.5  \\ 
HD19820 (2009 December 23)  &   4.70  $\pm$   0.10  &    115.0  $\pm$      0.6 &          -            &            -         &  4.61  $\pm$   0.07  &    114.3  $\pm$      0.4 &   4.06  $\pm$   0.08  &    114.0  $\pm$      0.5  \\
\hline
\end{longtable}
$^{\dagger}$ Degree of polarization and position angle values from \cite{1992AJ....104.1563S}.
\end{center}
\end{landscape}

\onecolumn 
\begin{longtable}{cccccccc}
\caption[R-band polarimetric and 2MASS data for the 127 stars]{R-band polarimetric and 2MASS data for the 127 stars.} \label{r-band-127} \\\hline 
\textbf{Star ID} & \textbf{R.A  ($\degr$)} & \textbf{DEC ($\degr$)} & \textbf{$P_{R}\pm \epsilon_{P}$}& \textbf{$\theta_{R}\pm\epsilon_{\theta}$}& \textbf{$J \pm \epsilon_{J}$}& \textbf{$H \pm \epsilon_{H}$}& \textbf{$Ks \pm \epsilon_{K}$} \\
\endfirsthead
\multicolumn{8}{c}%
{\tablename\ \thetable\ -- \textit{Continued from previous page}} \\
\hline
\textbf{Star ID} & \textbf{R.A  ($\degr$)} & \textbf{DEC ($\degr$)} & \textbf{$P_{R}\pm \epsilon$}& \textbf{$\theta_{R}\pm\epsilon$}& \textbf{$J \pm \epsilon$}& \textbf{$H \pm \epsilon$}& \textbf{$Ks \pm \epsilon$} \\
\hline
\endhead
\hline \multicolumn{8}{r}{\textit{Continued on next page}} \\
\endfoot
\hline
\endlastfoot

          &  (2000J)          & (2000J)         &  (\%)         &       ($\degr$)            & (mag)        &   (mag)           &   (mag)           \\
 (1)      &  (2)              &   (3)           &   (4)               &  (5)                       &   (6)        &  (7)              &  (8)              \\
\hline
       1  & 91.625190 &  19.376682 &   1.2 $\pm$  0.1 &  172 $\pm$  2 &    9.62 $\pm$   0.02 &    9.53 $\pm$   0.02 &    9.52 $\pm$   0.02 \\
       2  & 91.646004 &  19.428179 &   3.9 $\pm$  0.2 &  172 $\pm$  2 &    9.93 $\pm$   0.02 &    9.31 $\pm$   0.02 &    9.15 $\pm$   0.02 \\
       3  & 91.653043 &  19.370031 &   3.3 $\pm$  0.1 &  169 $\pm$  1 &    7.97 $\pm$   0.03 &    7.14 $\pm$   0.02 &    6.93 $\pm$   0.03 \\
       4  & 91.669906 &  19.410091 &   2.9 $\pm$  0.2 &  171 $\pm$  2 &    8.81 $\pm$   0.02 &    7.96 $\pm$   0.02 &    7.73 $\pm$   0.03 \\
       5  & 91.679249 &  19.535978 &   3.9 $\pm$  0.2 &  160 $\pm$  2 &   11.25 $\pm$   0.02 &   10.97 $\pm$   0.02 &   10.78 $\pm$   0.02 \\
       6  & 91.688594 &  19.496136 &   1.2 $\pm$  0.2 &  161 $\pm$  7 &   11.94 $\pm$   0.02 &   11.70 $\pm$   0.02 &   11.65 $\pm$   0.02 \\
\end{longtable}

$^\dagger$ H$\alpha$ emission line sources.

\small

\begin{center}
\begin{longtable}{ccccccccccccc}
\caption[R-band polarimetric and 2MASS data for the 127 stars]{Polarization and Serkowski fit parameters for the 42 stars with V$(RI)_{c}$ pass-band data} \label{vri_pol42} \\\hline 

\textbf{Star ID} & \textbf{R.A  ($\degr$)} & \textbf{DEC ($\degr$)} & \textbf{$P_{max}\pm \epsilon_{P}$}& \textbf{$\lambda_{max}\pm\epsilon$}& \textbf{$\sigma_{1}$}& \textbf{$\overline\epsilon$}& \textbf{$P_{V}\pm\epsilon$}& \textbf{$P_{Rc}\pm\epsilon$}& \textbf{$P_{Ic}\pm\epsilon $}& \textbf{$\theta_{V}\pm\epsilon$}& \textbf{$\theta_{Rc}\pm\epsilon$}& \textbf{$\theta_{Ic}\pm\epsilon$}\\\hline
\endfirsthead
\multicolumn{13}{c}%
{\tablename\ \thetable\ -- \textit{Continued from previous page}} \\
\hline
\textbf{Star ID} & \textbf{R.A  ($\degr$)} & \textbf{DEC ($\degr$)} & \textbf{$P_{max}\pm \epsilon_{P}$}& \textbf{$\lambda_{max}\pm\epsilon$}& \textbf{$\sigma_{1}$}& \textbf{$\overline\epsilon$}& \textbf{$P_{V}\pm\epsilon$}& \textbf{$P_{Rc}\pm\epsilon$}& \textbf{$P_{Ic}\pm\epsilon $}& \textbf{$\theta_{V}\pm\epsilon$}& \textbf{$\theta_{Rc}\pm\epsilon$}& \textbf{$\theta_{Ic}\pm\epsilon$}\\
\hline
\endhead
\hline \multicolumn{13}{r}{\textit{Continued on next page}} \\
\endfoot
\hline
\endlastfoot
      &   (2000J)  & (2000J) &  (per cent) & ($\mu$m) &  &  & (per cent) & (per cent) & (per cent) & ($\degr$) & ($\degr$) 
      & ($\degr$) \\
 (1)   &  (2)     &   (3)    &   (4)  &  (5)   &   (6)   &  (7)      &  (8)        &   (9)  &  (10)  &  (11) &  (12) & (13)  \\\hline
   5 &    91.679249 &  19.535978 &   3.9 $\pm$   0.2 &  0.62 $\pm$  0.07 &   0.2 &   0.9 &   3.8 $\pm$  0.3 &   3.9 $\pm$  0.2 &   3.6 $\pm$  0.2 &  163 $\pm$   2 &  160 $\pm$   2 &  160 $\pm$   2  \\
   6 &    91.688594 &  19.496136 &   1.3 $\pm$   0.3 &  0.51 $\pm$  0.18 &   0.3 &   0.4 &   1.3 $\pm$  0.3 &   1.2 $\pm$  0.2 &   1.0 $\pm$  0.3 &  167 $\pm$   6 &  161 $\pm$   7 &  165 $\pm$   4 \\
   7 &    91.689250 &  19.490425 &   4.2 $\pm$   0.3 &  0.54 $\pm$  0.06 &   1.2 &   0.5 &   4.4 $\pm$  0.3 &   3.8 $\pm$  0.2 &   3.7 $\pm$  0.2 &  183 $\pm$   2 &  183 $\pm$   2 &  185 $\pm$   2  \\
   8 &    91.694681 &  19.522718 &   1.4 $\pm$   0.2 &  0.53 $\pm$  0.13 &   0.5 &   1.1 &   1.4 $\pm$  0.2 &   1.4 $\pm$  0.2 &   1.1 $\pm$  0.2 &  174 $\pm$   4 &  168 $\pm$   5 &  166 $\pm$   3 \\
  20 &    91.733344 &  19.717730 &   2.7 $\pm$   0.3 &  0.54 $\pm$  0.10 &   3.0 &   1.7 &   2.2 $\pm$  0.4 &   3.1 $\pm$  0.2 &   1.9 $\pm$  0.2 &  162 $\pm$   4 &  169 $\pm$   4 &  151 $\pm$   2 \\
  24 &    91.755495 &  19.662973 &   4.1 $\pm$   0.8 &  1.09 $\pm$  0.21 &   0.2 &   0.8 &   2.2 $\pm$  0.4 &   3.2 $\pm$  0.3 &   3.7 $\pm$  0.3 &  160 $\pm$   6 &  157 $\pm$   2 &  165 $\pm$   2  \\
\hline \hline     
\end{longtable}
\end{center}

\begin{landscape} 
\begin{center}
\begin{longtable}{ccccccccccccccc}
\caption[]{Polarization and Serkowski fit parameters for the 15 stars with $B$, $V$ $(RI)_{c}$ pass-band data} \label{bvri_pol15} \\\hline 

\textbf{Star ID} & \textbf{R.A  ($\degr$)} & \textbf{DEC ($\degr$)} & \textbf{$P_{max}\pm \epsilon_{P}$}& \textbf{$\lambda_{max}\pm\epsilon$}& \textbf{$\sigma_{1}$}& \textbf{$\overline\epsilon$}& \textbf{$P_{B}\pm\epsilon$}& \textbf{$P_{V}\pm\epsilon$}& \textbf{$P_{Rc}\pm\epsilon$}& \textbf{$P_{Ic}\pm\epsilon $}&\textbf{$\theta_{B}\pm\epsilon$}& \textbf{$\theta_{V}\pm\epsilon$}& \textbf{$\theta_{Rc}\pm\epsilon$}& \textbf{$\theta_{Ic}\pm\epsilon$}\\
\endfirsthead
\multicolumn{15}{c}%
{\tablename\ \thetable\ -- \textit{Continued from previous page}} \\
\hline
\textbf{Star ID} & \textbf{R.A  ($\degr$)} & \textbf{DEC ($\degr$)} & \textbf{$P_{max}\pm \epsilon_{P}$}& \textbf{$\lambda_{max}\pm\epsilon$}& \textbf{$\sigma_{1}$}& \textbf{$\overline\epsilon$}&\textbf{$\theta_{B}\pm\epsilon$}& \textbf{$P_{V}\pm\epsilon$}& \textbf{$P_{Rc}\pm\epsilon$}& \textbf{$P_{Ic}\pm\epsilon $}&\textbf{$\theta_{V}\pm\epsilon$}& \textbf{$\theta_{V}\pm\epsilon$}& \textbf{$\theta_{Rc}\pm\epsilon$}& \textbf{$\theta_{Ic}\pm\epsilon$}\\
\hline
\endhead
\hline \multicolumn{15}{r}{\textit{Continued on next page}} \\
\endfoot
\hline
\endlastfoot
      &   (2000J)  & (2000J) &  (per cent) & ($\mu$m) &  &  & (per cent) & (per cent) & (per cent) & (per cent) & ($\degr$) & ($\degr$)& ($\degr$) 
      & ($\degr$)  \\
 (1)   &  (2)     &   (3)    &   (4)  &  (5)   &   (6)   &  (7)      &  (8)        &   (9)  &  (10)  &  (11) &  (12) & (13) & (14) & (15) \\\hline

  14 &   91.720034 &  19.573496 &   0.2 $\pm$   0.1 &  0.73 $\pm$  0.19 &   0.5 &   2.2 &   0.1 $\pm$  0.1 &   0.2 $\pm$  0.1 &   0.2 $\pm$  0.1 &   0.2 $\pm$  0.1 &  179 $\pm$  15 &  181 $\pm$   9 &  187 $\pm$   6 &  108 $\pm$   8 \\
  19 &   91.732852 &  19.580721 &   1.8 $\pm$   0.1 &  0.48 $\pm$  0.03 &   1.4 &   2.3 &   1.7 $\pm$  0.1 &   1.9 $\pm$  0.1 &   1.6 $\pm$  0.1 &   1.4 $\pm$  0.1 &  178 $\pm$   1 &  174 $\pm$   1 &  179 $\pm$   1 &  173 $\pm$   1 \\
  23 &   91.753682 &  19.573689 &   1.7 $\pm$   0.1 &  0.33 $\pm$  0.02 &  16.9 &  17.8 &   1.7 $\pm$  0.1 &   1.3 $\pm$  0.1 &   1.1 $\pm$  0.1 &   1.4 $\pm$  0.1 &  177 $\pm$   1 &  168 $\pm$   1 &  181 $\pm$   1 &  170 $\pm$   1  \\
  30 &   91.770629 &  19.585886 &   4.1 $\pm$   0.1 &  0.55 $\pm$  0.02 &   2.3 &   7.8 &   3.7 $\pm$  0.1 &   4.3 $\pm$  0.1 &   4.1 $\pm$  0.1 &   3.5 $\pm$  0.1 &  184 $\pm$   1 &  173 $\pm$   1 &  182 $\pm$   1 &  172 $\pm$   1 \\
  48$^\dagger$ &   91.816043 &  19.500032 &   1.1 $\pm$   0.1 &  0.63 $\pm$  0.08 &   1.4 &   3.2 &   0.7 $\pm$  0.2 &   1.1 $\pm$  0.1 &   1.2 $\pm$  0.1 &   1.0 $\pm$  0.1 &  192 $\pm$   7 &  159 $\pm$   3 &  154 $\pm$   2 &  149 $\pm$   2  \\
  83 &   91.915392 &  19.558058 &   2.9 $\pm$   0.1 &  0.66 $\pm$  0.05 &   2.1 &   1.4 &   2.3 $\pm$  0.3 &   3.0 $\pm$  0.2 &   2.5 $\pm$  0.2 &   3.1 $\pm$  0.2 &  176 $\pm$   3 &  179 $\pm$   2 &  172 $\pm$   2 &  175 $\pm$   2  \\
\hline \hline
\end{longtable}
\end{center}
\end{landscape}
\begin{table}
\caption{The details of the fields used for estimating distance to L1570. The Galactic coordinates given here are the central coordinates of the fields. \label{tab:fields}}
\begin{tabular}{lcclll}\hline
Field&$ l $&$ b $&Cloud&Dark &$V_{LSR}$\\
Id    &&&Name$ ^{\dagger} $&clouds&\\
&($ ^{\circ} $)&($ ^{\circ} $)&&&($km~s^{-1}$)\\
\hline
1     &190.8059  & -0.3097 &-              	&L1570	&$-$0.5$^{\ddagger} $\\
2     &191.5789  & -0.7500 &191.6-00.8	&L1576	&$+$0.3\\
3     &192.0724  & -0.8478 &192.1-00.9	&L1578	&$+$0.2\\
4     &185.1870  & -2.1776 &185.3-02.1	&-         	&$-$0.2\\
5     &185.6635  & -2.0188 &185.3-02.1	&-         	&$-$0.2\\
6     &189.0888  & -1.8889 &188.9-02.0	&-         	&$-$0.8\\
\hline
\end{tabular}

$ ^{\dagger} $ Cloud names are taken from \citet{1998ApJS..117..387K}.\\
$^{\ddagger} $ $V_{LSR}$ value is taken from \citet{1988ApJS...68..257C}.
\end{table}
\small
\begin{center}
\begin{longtable}{cccccccccc}
\caption[Photometric data of 144 stars with good signal-noise-ratio]{Photometric data of 144 stars with good signal-noise-ratio (with $BVRIJHK$-photometric errors \textless 0.1 mag)}\label{photdata} \\ \hline 

\textbf{Star ID} & \textbf{R.A  ($\degr$)} & \textbf{DEC ($\degr$)} & \textbf{$B \pm \epsilon_{B}$} & \textbf{$V \pm \epsilon_{V}$} & \textbf{$R \pm \epsilon_{R}$} &  \textbf{$I \pm \epsilon_{I}$} &  \textbf{$J \pm \epsilon_{J}$} &  \textbf{$H \pm \epsilon_{H}$} &  \textbf{$K \pm \epsilon_{I}$}   \\
\endfirsthead
\multicolumn{10}{c}%
{\tablename\ \thetable\ -- \textit{Continued from previous page}} \\
\hline

\hline\textbf{Star ID} & \textbf{R.A  ($\degr$)} & \textbf{DEC ($\degr$)} & \textbf{$B \pm \epsilon_{B}$} & \textbf{$V \pm\epsilon_{V}$}   & \textbf{$R \pm \epsilon_{R}$} &  \textbf{$I \pm \epsilon_{I}$} &  \textbf{$J \pm \epsilon_{J}$} &  \textbf{$H \pm \epsilon_{H}$} &  \textbf{$K \pm \epsilon_{I}$} \\\hline
\endhead
\multicolumn{10}{r}%
{\tablename\ \thetable\ -- \textit{Continued on next page}} \\
\endfoot
\hline
\endlastfoot

    &   (2000J)  & (2000J) &  (mag) & (mag) & (mag) & (mag) & (mag)  & (mag)  & (mag)  \\
 (1)   &  (2)     &   (3)    &   (4)  &  (5)   &   (6)   &  (7)      &  (8)        &   (9)  &  (10)   \\\hline

     1 &   91.750253 &   19.466185 &   15.94 $\pm$    0.01 &   15.06 $\pm$    0.00 &   14.55 $\pm$    0.01 &   13.99  $\pm$   0.01 &   13.20 $\pm$    0.02 &   12.85 $\pm$    0.02 &   12.73 $\pm$    0.02 \\
      2 &   91.755159 &   19.455719 &   16.92 $\pm$    0.02 &   15.96 $\pm$    0.01 &   15.45 $\pm$    0.02 &   14.88  $\pm$   0.02 &   14.06 $\pm$    0.03 &   13.66 $\pm$    0.03 &   13.47 $\pm$    0.03 \\
      3 &   91.757644 &   19.470079 &   15.80 $\pm$    0.01 &   15.02 $\pm$    0.00 &   14.58 $\pm$    0.01 &   14.08  $\pm$   0.01 &   13.34 $\pm$    0.02 &   13.10 $\pm$    0.02 &   12.93 $\pm$    0.02 \\
      4 &   91.758920 &   19.492409 &   17.92 $\pm$    0.03 &   16.74 $\pm$    0.01 &   16.02 $\pm$    0.02 &   15.25  $\pm$   0.02 &   14.16 $\pm$    0.04 &   13.75 $\pm$    0.04 &   13.53 $\pm$    0.04 \\
      5 &   91.762366 &   19.474054 &   18.62 $\pm$    0.05 &   17.35 $\pm$    0.01 &   16.60 $\pm$    0.02 &   15.84  $\pm$   0.03 &   14.83 $\pm$    0.03 &   14.25 $\pm$    0.04 &   14.01 $\pm$    0.04 \\
      6 &   91.762867 &   19.487282 &   17.64 $\pm$    0.02 &   16.50 $\pm$    0.01 &   15.82 $\pm$    0.01 &   15.16  $\pm$   0.01 &   14.17 $\pm$    0.03 &   13.53 $\pm$    0.03 &   13.41 $\pm$    0.03 \\
\end{longtable}
\end{center}
$^\dagger$ H$\alpha$ emission line sources. \\
$^\star$ Unreddened M-type dwarfs identified using $(V-I)$ vs $(B-V)$ colour-colour diagram. \\
Note: The star ids mentioned in brackets are the stars those with polarimetric observations mentioned in the Table \ref{r-band-127}.  \\

\begin{table}
\begin{small}
\caption{Slopes of the color-color combination of the stars
distributed towards L1570.}\label{slopes_TCDs}
\begin{tabular}{llll}\hline \hline

color-color & $m_{cloud}$ &  $m_{cloud}$ & $m_{normal}$ \\
combination &  (using 26 stars) & (using 135 stars) &  \\
\hline
$(B-V)/(V-I)$  &   -1.30 $\pm$ 0.01 & -1.36$\pm$0.01   & -1.10   \\
$(B-V)/(V-J)$  &   -2.24 $\pm$ 0.03 & -2.34$\pm$0.02   & -1.96   \\
$(B-V)/(V-H)$  &   -2.68 $\pm$ 0.03 & -2.77$\pm$0.02   & -2.42   \\
$(B-V)/(V-K)$  &   -2.86 $\pm$ 0.02 & -2.94$\pm$0.02   & -2.60   \\
\hline \hline
\end{tabular}
\end{small}
\end{table}

                                                                                                                                                                                                         
\end{document}